\newcommand{\avg}[1]{\left< #1 \right>} 
\newcommand*{\ditto}{\raisebox{-0.5ex}{\ttfamily"}}

\newcommand*\diff{\mathop{}\!\mathrm{d}}

\documentclass[manuscript]{aastex62}

\usepackage{amsbsy}
\usepackage{amsfonts}
\usepackage{amssymb}
\usepackage{graphicx}
\usepackage{amsmath}
\usepackage{graphicx}
\usepackage{graphics}
\usepackage{bm}
\usepackage{natbib}
\usepackage{array,longtable}
\usepackage{subfigure}
\usepackage{lipsum}

\graphicspath{{./}{figures/}}

\begin{document}
\title{Laboratory Calibrations of Fe~\textsc{xii}--\textsc{xiv} Line-Intensity Ratios for Electron Density Diagnostics.}

\correspondingauthor{Michael Hahn}
\email{mhahn@astro.columbia.edu}

\author{Thusitha Arthanayaka}
\affiliation{Columbia Astrophysics Laboratory, Columbia University, 550 West 120th Street, New York, NY 10027 USA}

\author{Peter Beiersdorfer}
\affiliation{Lawrence Livermore National Laboratory, Livermore, CA 94550, USA}

\author{Gregory V. Brown}
\affiliation{Lawrence Livermore National Laboratory, Livermore, CA 94550, USA}

\author{Ming Feng Gu}
\affiliation{Space Science Laboratory, University of California, Berkeley, CA 94720, USA}

\author{Michael Hahn}
\affiliation{Columbia Astrophysics Laboratory, Columbia University, 550 West 120th Street, New York, NY 10027 USA}

\author{Natalie Hell}
\affiliation{Lawrence Livermore National Laboratory, Livermore, CA 94550, USA}

\author{Tom Lockard}
\affiliation{Lawrence Livermore National Laboratory, Livermore, CA 94550, USA}

\author{Daniel W. Savin}
\affiliation{Columbia Astrophysics Laboratory, Columbia University, 550 West 120th Street, New York, NY 10027 USA}

\begin{abstract}

We have used an electron beam ion trap to measure electron-density-diagnostic line-intensity ratios for extreme ultraviolet lines from Fe~\textsc{xii}, \textsc{xiii}, and \textsc{xiv} at wavelengths of $\approx185$--205 and 255--276~\AA. These ratios can be used as density diagnostics for astrophysical spectra and are especially relevant to solar physics. We found that density diagnostics using the Fe~\textsc{xiii} 196.53/202.04 and the Fe~\textsc{xiv} 264.79/274.21 and 270.52A/274.21 line ratios are reliable using the atomic data calculated with the Flexible Atomic Code. On the other hand, we found a large discrepancy between the FAC theory and experiment for the commonly used Fe~\textsc{xii} (186.85 + 186.88)/195.12 line ratio. These FAC theory calculations give similar results to the data tabulated in CHIANTI, which are commonly used to analyze solar observations. Our results suggest that the discrepancies seen between solar coronal density measurements using the Fe~\textsc{xii} (186.85 + 186.88)/195.12 and Fe~\textsc{xiii} 196.54/202.04 line ratios are likely due to issues with the atomic calculations for Fe~\textsc{xii}. 

\end{abstract}


\section{Introduction}\label{sec:intro}

Reliable measurements of electron density, $n_{\mathrm{e}}$, are needed to understand many astrophysical systems. For example, in solar physics such measurements are necessary to quantify heating rates, infer wave properties, model explosive events such as flares, and understand the general structure of the solar atmosphere. Electron densities can be measured spectroscopically by taking the intensity ratio of two spectral lines, at least one of which is density sensitive \citep{Phillips:book}. The line pairs are chosen to come from the same ion species so that their ratio is only sensitive to density and insensitive to temperature, charge state abundance, or elemental abundance. Such line pairs are referred to as density diagnostics. 

Density sensitivity can arise in several ways. For the lines of interest here, both transitions forming the lines are allowed, but the upper level of one of the lines is populated preferentially from a metastable lower level. The population of that metastable level is sensitive to density because its population depends mainly on collisional excitation into that level from lower levels and collisional excitation or de-excitation out of that level, with radiative decay being important only at low densities. The density-insensitive member of the line pair is populated mainly by collisions from the ground level and depopulated by radiative decays.  

Theoretical calculations to predict density-sensitive line ratios require accurate modeling of many aspects of the atomic physics. The atomic model must include reliable excitation cross sections for populating both the metastable levels and the upper levels that form the lines, radiative decay rates from all those levels, and collisional de-excitation rates. The atomic model must also be large enough to capture all the important cascade effects from higher levels, which requires accurate collision data and radiative lifetimes for all those levels. With so many interrelated aspects of the atomic model contributing to the density sensitivity, many sources of uncertainty can combine to lead to significant errors. 

Recent studies have shown that the calculations that underlie density diagnostics do indeed have large systematic uncertainties. \citet{Young:AA:2009} inferred the density of a solar active region using density diagnostics from Fe~\textsc{xii} and Fe~\textsc{xiii}. They found that the inferred densities differed from one another by up to a factor of $\sim 10$ in some cases, with typical discrepancies of a factor of $\sim 3$. These two ions form at very similar temperatures and so are expected to probe essentially the same parcel of gas. Hence, these discrepancies are most likely due to inaccurate atomic data. This explanation is supported by the work of \citet{Watanabe:ApJ:2009}, who compared Fe~\textsc{xiii} density diagnostics using several different atomic data sources and found that the inferred densities could vary by a factor of four depending on which set of atomic data was used to interpret the line ratio. We note that Fe~\textsc{xii} and \textsc{xiii} are not unique in having uncertain density diagnostics. They have been a focus of study because they are commonly used lines for solar physics, not because there is any inherent reason to expect that their diagnostics are more inaccurate than those from any other system.

Experiments are needed in order to obtain empirical calibrations of density diagnostics and to provide benchmarks for improving the atomic calculations. Electron beam ion traps (EBITs) have been used previously to study density diagnostic line ratios \citep{Chen:ApJ:2004, Yamamoto:ApJ:2008, Liang:ApJ:2009a, Liang:ApJ:2009, Nakamura:ApJ:2011, Shimizu:AA:2017}. 
Here we present EBIT measurements of density diagnostic line ratios for Fe~\textsc{xii}, \textsc{xiii} and \textsc{xiv}. The measurements focus on lines that are commonly measured in solar physics \citep{Young:PASJ:2007, Watanabe:ApJ:2009}. Many of these lines are observed by the Extreme Ultraviolet Imaging Spectrometer \citep[EIS;][]{Culhane:SolPhys:2007} on \textit{Hinode}. Our experiments were done using the EBIT-I device at the Lawrence Livermore National Laboratory (LLNL). 

\section{Experiment and Analysis} \label{sec:experiment}

Details of the experiment using EBIT-I have been described previously in \citet{Arthanayaka:RSI:2018}. Briefly, the ions were confined axially by a potential well created by applied voltages to a set of three copper drift tubes. An axial trapping potential of 400~V was used. The radial confinement was provided by the electrostatic attraction of the ions to the electron beam as well as by a 3~T axial magnetic field. Iron was introduced into the trap as a gas of iron pentacarbonyl [$\mathrm{Fe(CO)_{5}}$] using a collimated continuous ballistic gas injection system. The gas was dissociated and ionized by the electron beam. The resulting trapped Fe ions were further ionized and excited by collisions with the electron beam.

The density in EBIT-I can be varied by changing the electron beam energy or current. The electron energy, $E_{\mathrm{e}}$, in the trapping region can be controlled by varying the voltages of the three drift tubes \citep[see][]{Arthanayaka:RSI:2018}. For our experiments, the two beam energies used were nominally $E_{\mathrm{e}}=395$~eV and 475~eV, taking into account the $\approx 20$~eV reduction due to space-charge effects in the electron beam \citep{Vogel1990}. The experiments were carried out using beam currents ranging from $I_{\mathrm{e}}  = 1$--$9$~mA. 

\subsection{Spectra and Line Fitting} \label{subsec:specline}

The radius of the electron beam, $r_{\mathrm{e}}$, 
and the emission spectra of the ions were measured simultaneously using a high-resolution grazing incidence grating spectrometer \citep[HiGGS;][]{Beiersdorfer:RSI:2014}. The spectrometer has a grating with 2400 lines/mm, a radius of curvature of $R= 44.3$~m, and a liquid-nitrogen-cooled charge-coupled device (CCD). No entrance slit was used in the setup. The allowed spectral lines observed in the EUV are emitted from within the $\sim 60$~$\mu\mathrm{m}$ wide electron beam \citep{Marrs:PhysScr:1995, Utter:NIMA:1999}, thereby forming a narrow emission source that effectively serves as a slit \citep[see e.g.,][]{Liang:ApJ:2009, Liang:ApJ:2009a}.

We carried out EUV measurements in the ranges of $\approx 185$--$205$ and $\approx 255$--$276$~\AA. In order to accurately determine the intensity ratio of spectral lines, it is necessary to correctly identify the spectral lines and resolve any line blends. For our EUV measurements, we calibrated the wavelength range by using well known intense lines from ions of O and Fe. Wavelength listings from NIST \citep{NIST:2018} and CHIANTI \citep{Dere:AAS:1997, Landi:ApJ:2013, DelZanna:AA:2015}  as well as our previous line identification experiments \citep{Beiersdorfer:ApJ:2014, Trabert:ApJS:2014b} were used to identify these and the other lines in our measured spectra. However, because of the complexity of the spectrum in this wavelength range, not all the lines could be identified. The resolving power, $\lambda / \Delta \lambda$, of the HiGGS was in the range 2800--4000 for our experiments, where $\Delta \lambda$ is the measured line width and $\lambda$ is the line center. This is similar to that of spectrometers used for solar observations, such as EIS on \textit{Hinode}. Observed spectral lines may have blends from the same ion species or nearby charge states of the same ion. Since the Fe spectrum in our experiment is produced by a gas containing C and O atoms, blends with lines from those elements are also possible. 

\begin{figure}
	\centering \includegraphics[width=0.9\textwidth]{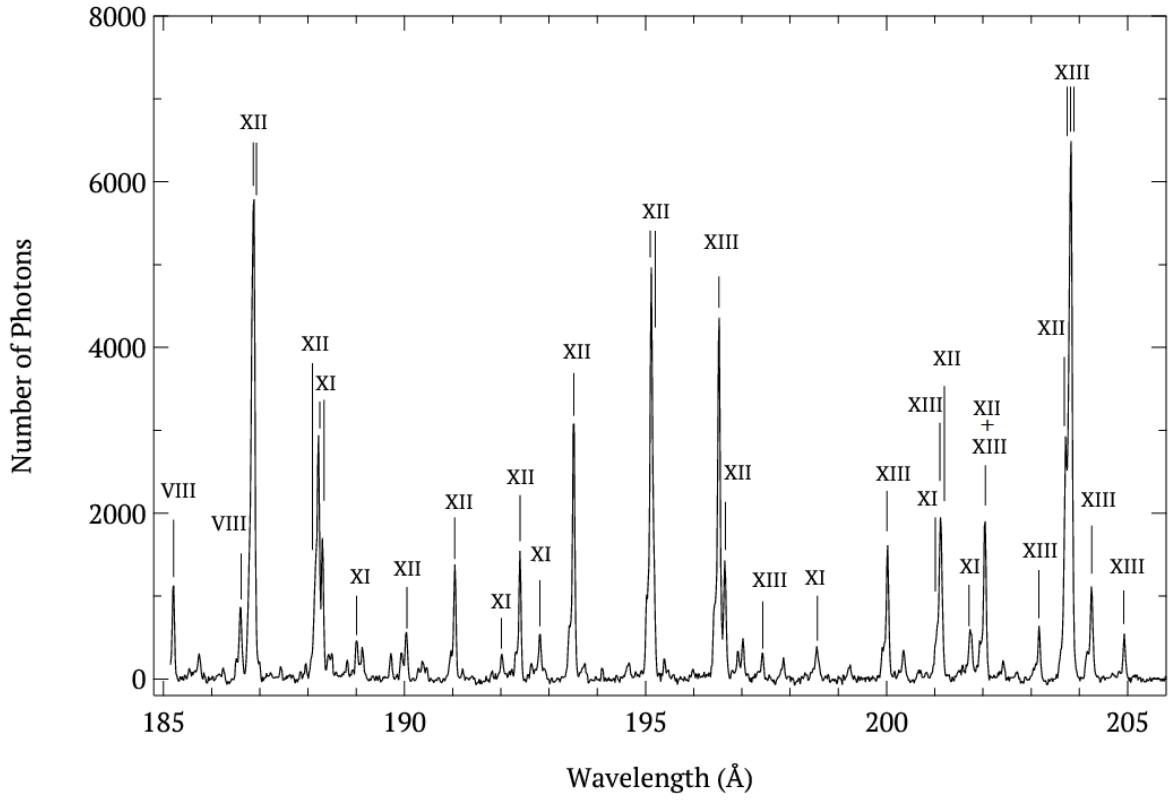}
	\caption{\label{fig:spec1} HiGGS spectrum obtained  for $E_{\mathrm{e}}$ = 395 eV and $I_{\mathrm{e}}= 7$~mA in the 185--205~\AA\ range. The Fe lines of interest for the various charge states are labeled with their corresponding spectroscopic Roman numerals. See Table~\ref{table:table1} for the detailed transition list.     
}

\end{figure}

Figure~\ref{fig:spec1} shows the spectrum obtained for $E_{\mathrm{e}} = 395$~eV and $I_{\mathrm{e}}=7$~mA in the 185--205~\AA\ spectral range. All the identified strong lines and other lines of interest are listed in Table~\ref{table:table1}. In this wavelength range the spectrum is dense and there were many blends. The measured full with at half maximum (FWHM) of the lines was $\Delta \lambda \approx 0.050$~\AA, which is set by the electron beam width. This corresponds to a resolving power of $\lambda/\Delta \lambda \approx 4000$. The magnification of the EUV spectrometer set up was $1.00 \pm 0.05$. The magnification is determined by the ratio of the distances of the grating from the detector and from the source. Here and throughout, all uncertainties are given for an estimated $1\sigma$ confidence level.

\begin{figure}
	\centering \includegraphics[width=0.9\textwidth]{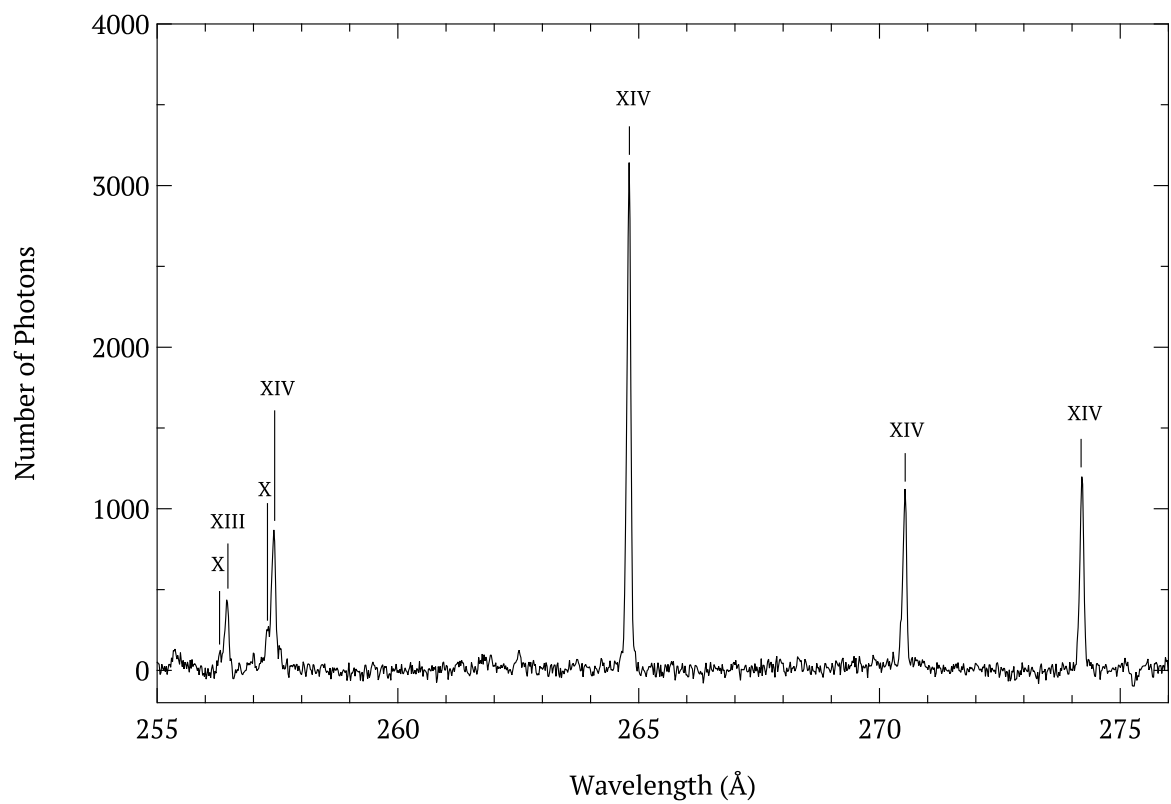}
	\caption{\label{fig:spec2} Same as Figure~\ref{fig:spec1}, but for the 255--276~\AA\ range. See Table \ref{table:table2} for the detailed transition list. 
}
\end{figure}

Figure~\ref{fig:spec2} shows the spectrum for the 255--276~\AA\ range at the same beam energy and current as Figure~\ref{fig:spec1}. The identified transitions of interest are listed in Table~\ref{table:table2}. In contrast to the shorter wavelength range, this region has spectral features that are less complicated and we were able to identify all of the important lines. 
The measured line width was $\approx 0.095$~\AA\ at a magnification of $1.03 \pm 0.05$ for this wavelength range. This corresponds to a resolving power of 2800. We did not perform experiments at the 475~eV beam energy for this spectral range, because our earlier experience with the shorter wavelength range showed no significant difference in the electron-density diagnostic calibrations as the electron beam energy was varied. 

We fit the observed spectra with Gaussian line profiles in order to extract the intensities of the various lines of interest, separate some line blends, and quantify the line widths. For these fits, the Gaussian widths were constrained to be equal, because those widths are set by the width of the electron beam and so should be the same for all unblended lines. The background level of the CCD was determined from the two-dimensional images, as described in \citet{Arthanayaka:RSI:2018}. This enabled us to set the background level to zero for the collapsed one-dimensional data on which we performed the fitting. In the least-squares fitting routine, we allowed the centroids to vary freely. The line fitting scheme for each set of lines in a given wavelength range was kept consistent for different beam currents. 

This fitting scheme is subject to some systematic error and other reasonable fitting schemes could have been chosen. For example, the background subtraction we performed is imperfect and so we could incorporate additional free parameters to account for the background level, such as a constant or a linear term. In principle, all the observed lines are expected to have the same width. However, unknown blends might cause minor broadening of the lines. To account for that, we could allow the widths of all the Gaussian components to be independent. We have tested the effects of these various fitting schemes and found that they cause the intensity ratio to vary by about $\pm 8\%$. 

\begin{figure}[ht!]
     \begin{center}
        \subfigure {%
            \label{fig:first}
            \includegraphics[width=0.4\textwidth]{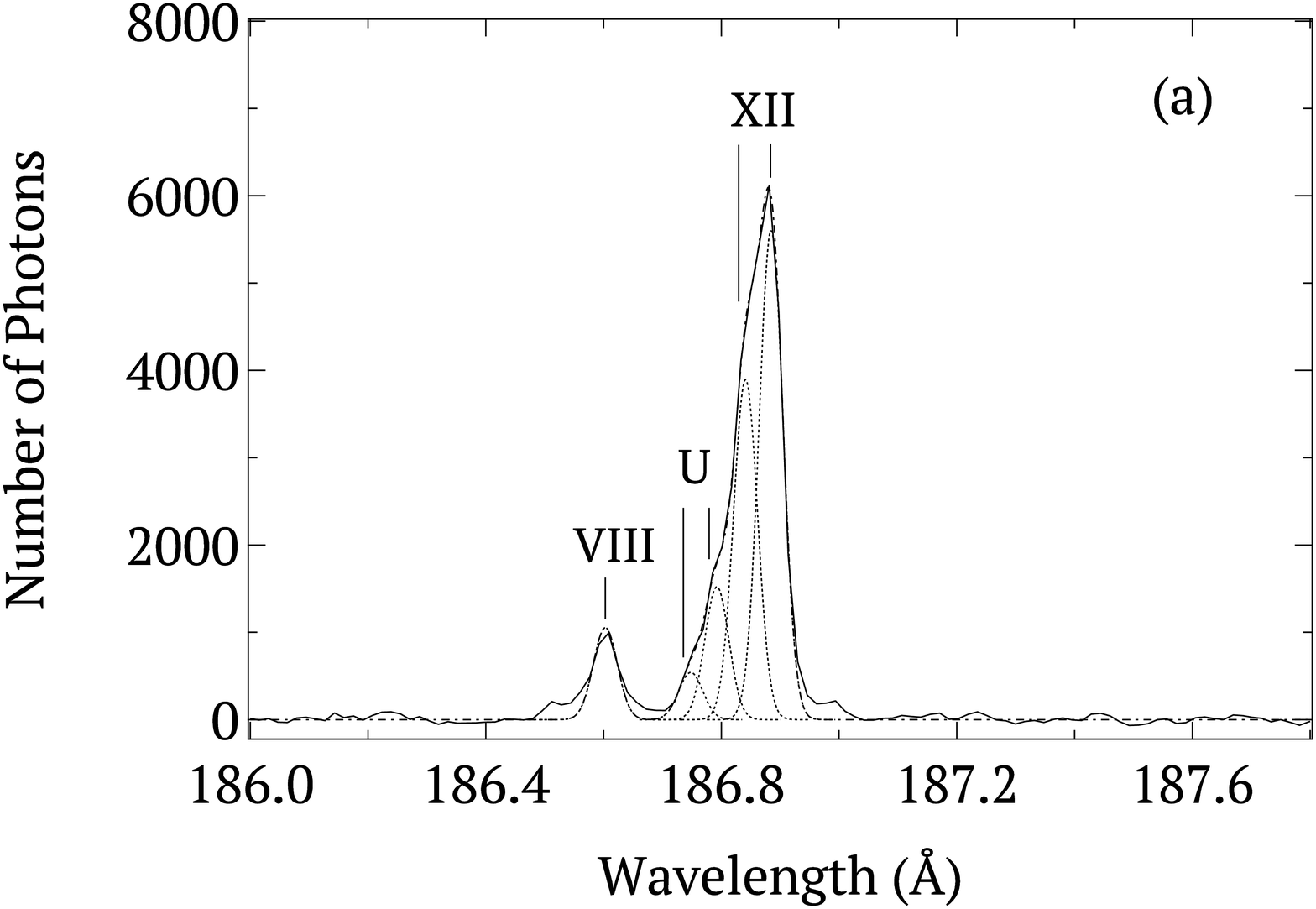}
            
        }%
        \subfigure{%
           \label{fig:second}
           \includegraphics[width=0.4\textwidth]{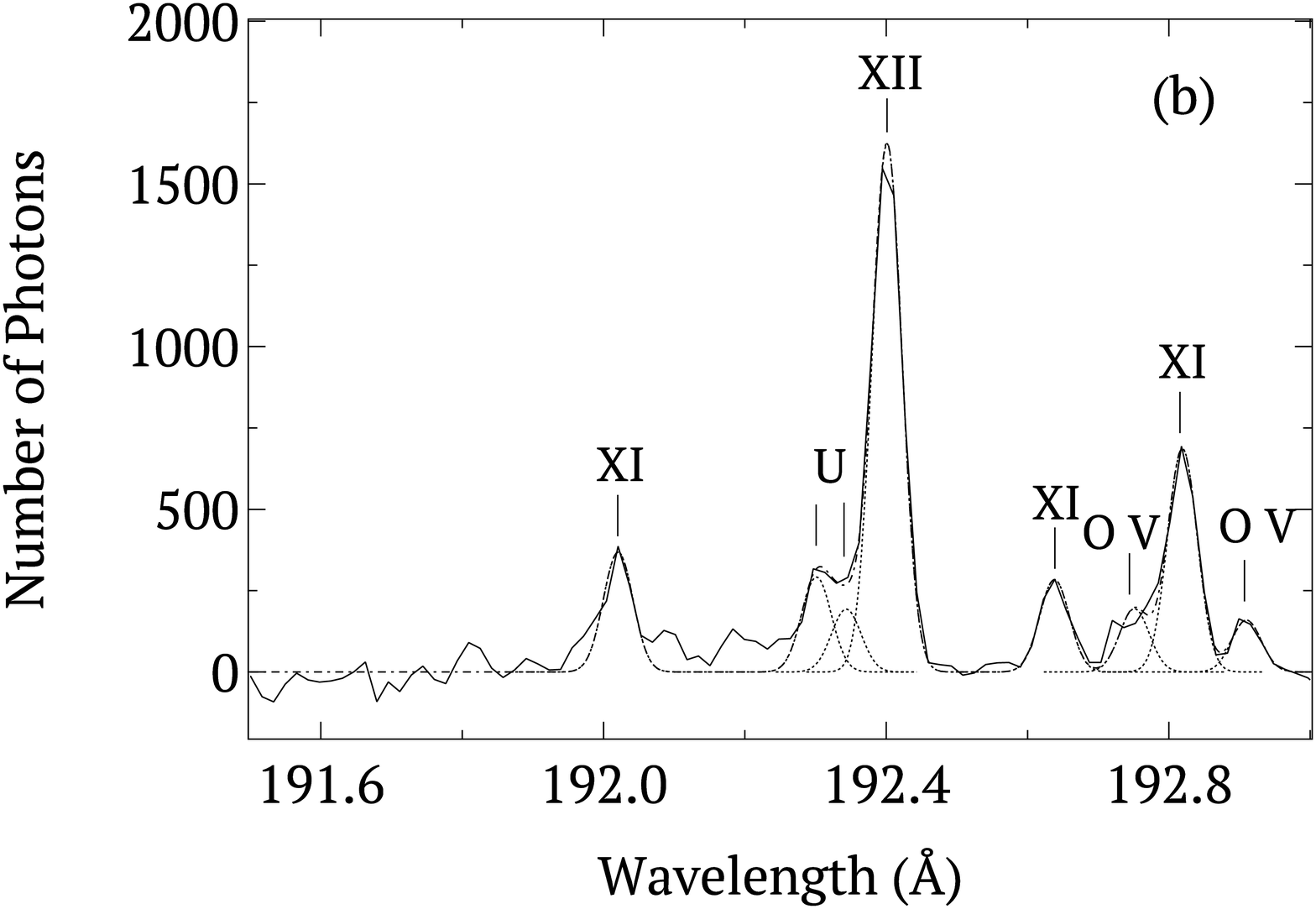}
       }\\ 
        \subfigure {%
            \label{fig:third}
            \includegraphics[width=0.4\textwidth]{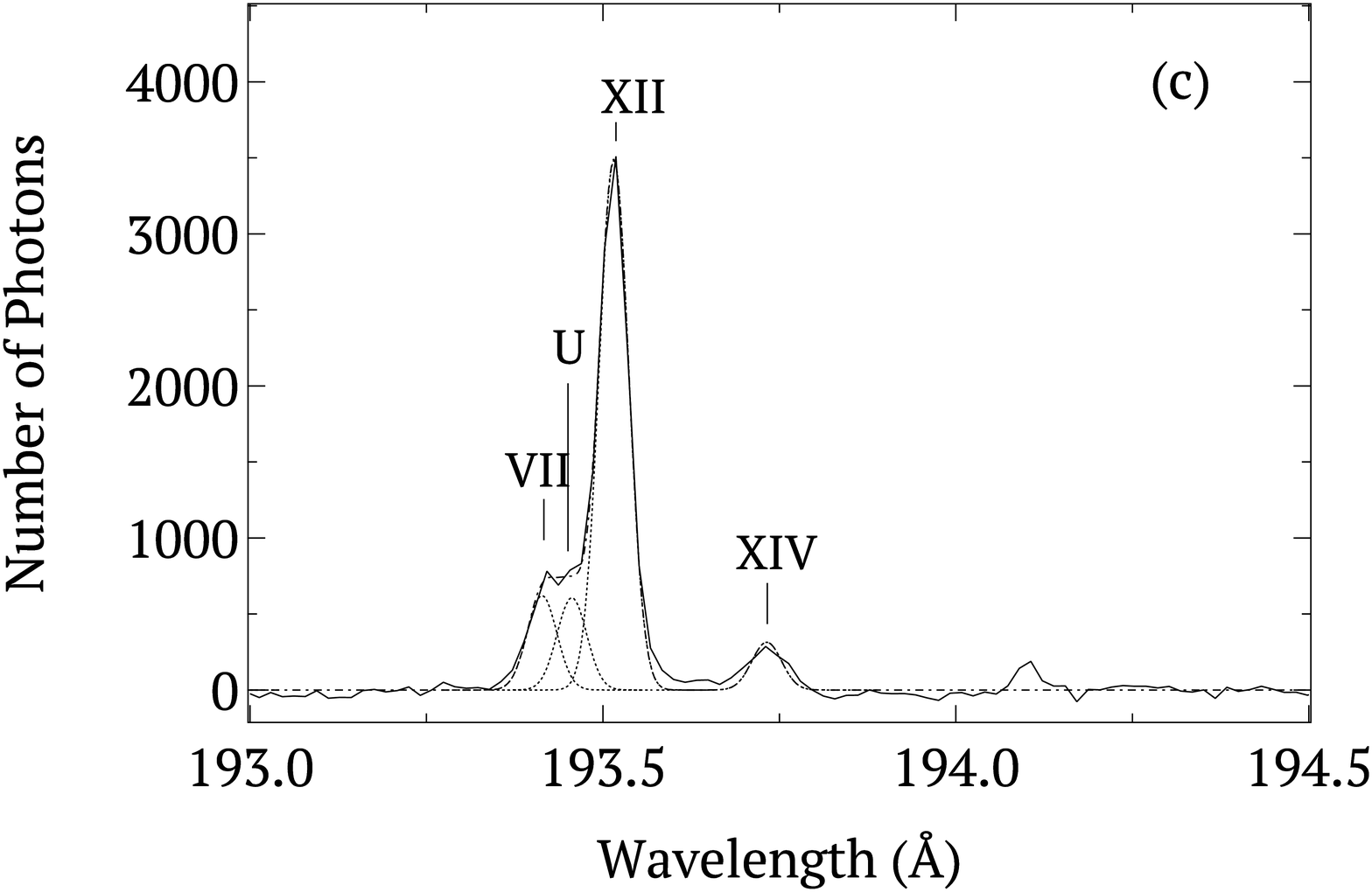}
        }%
        \subfigure{%
            \label{fig:fourth}
            \includegraphics[width=0.4\textwidth]{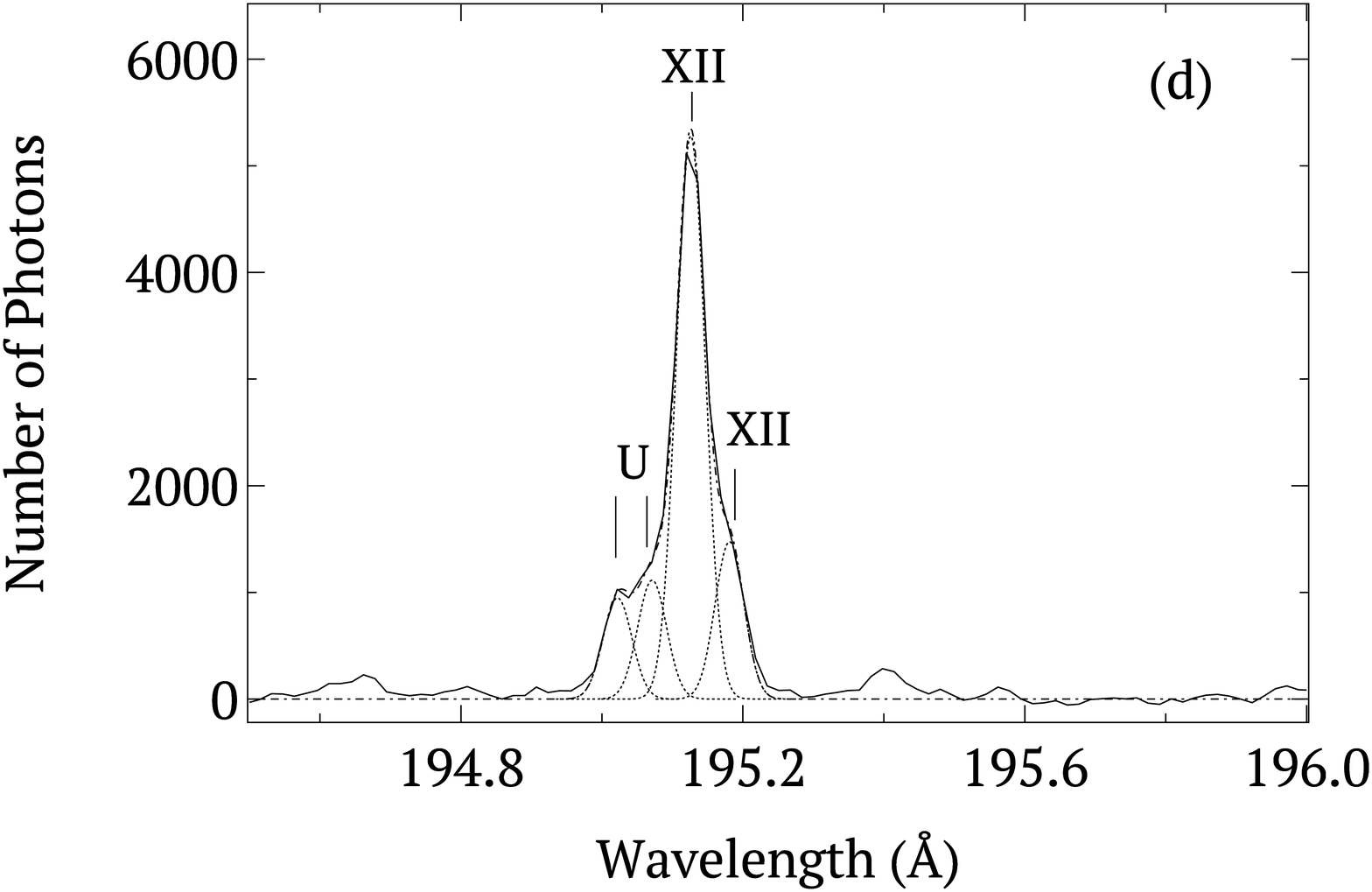}
        }%
        \\ 
        \subfigure {%
            \label{fig:fifth}
            \includegraphics[width=0.4\textwidth]{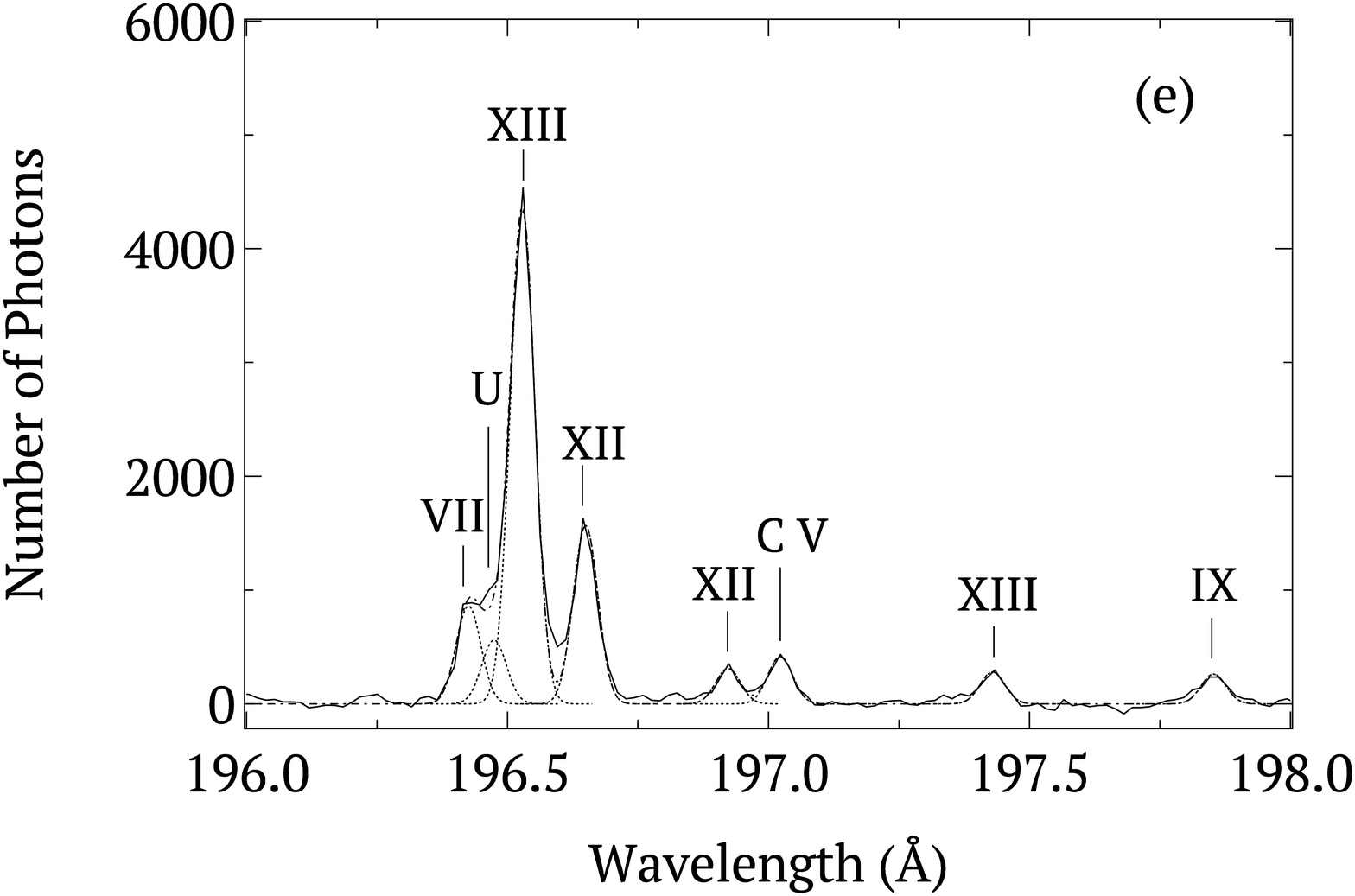}
        }%
        \subfigure {%
            \label{fig:sixth}
            \includegraphics[width=0.4\textwidth]{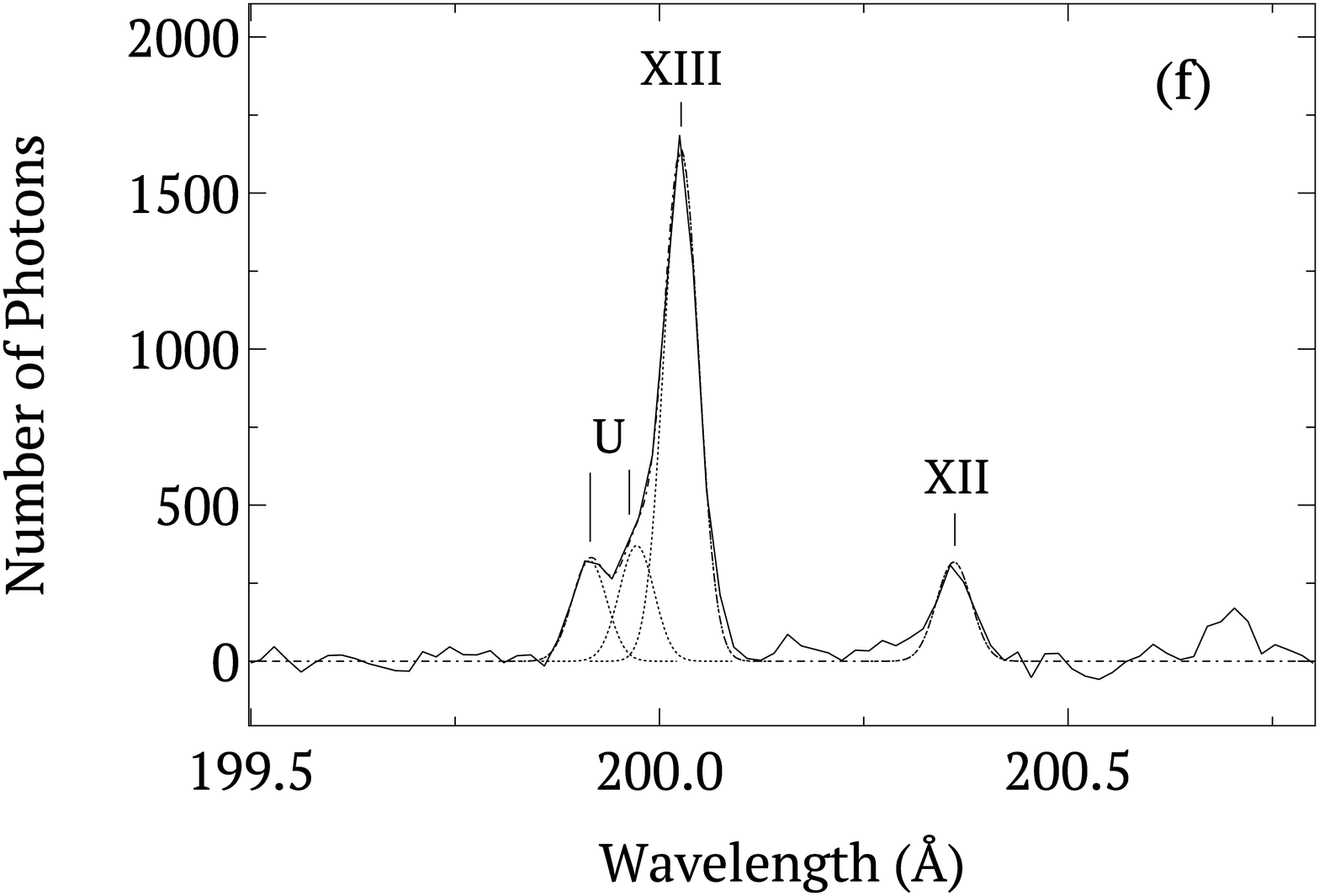}
        }\\ 
        \subfigure {%
            \label{fig:seventh}
            \includegraphics[width=0.4\textwidth]{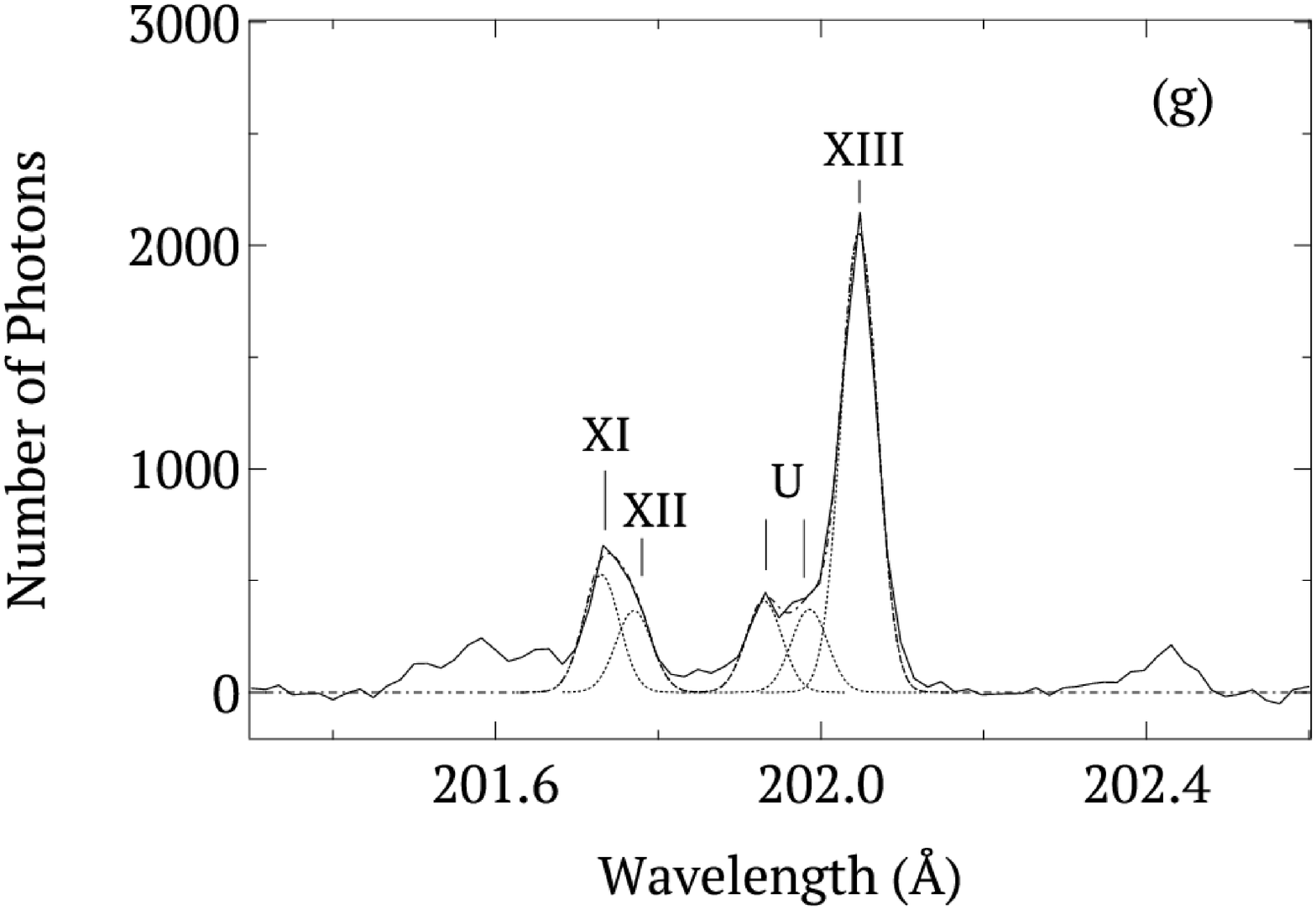}
        }%
        \subfigure {%
            \label{fig:eighth}
            \includegraphics[width=0.4\textwidth]{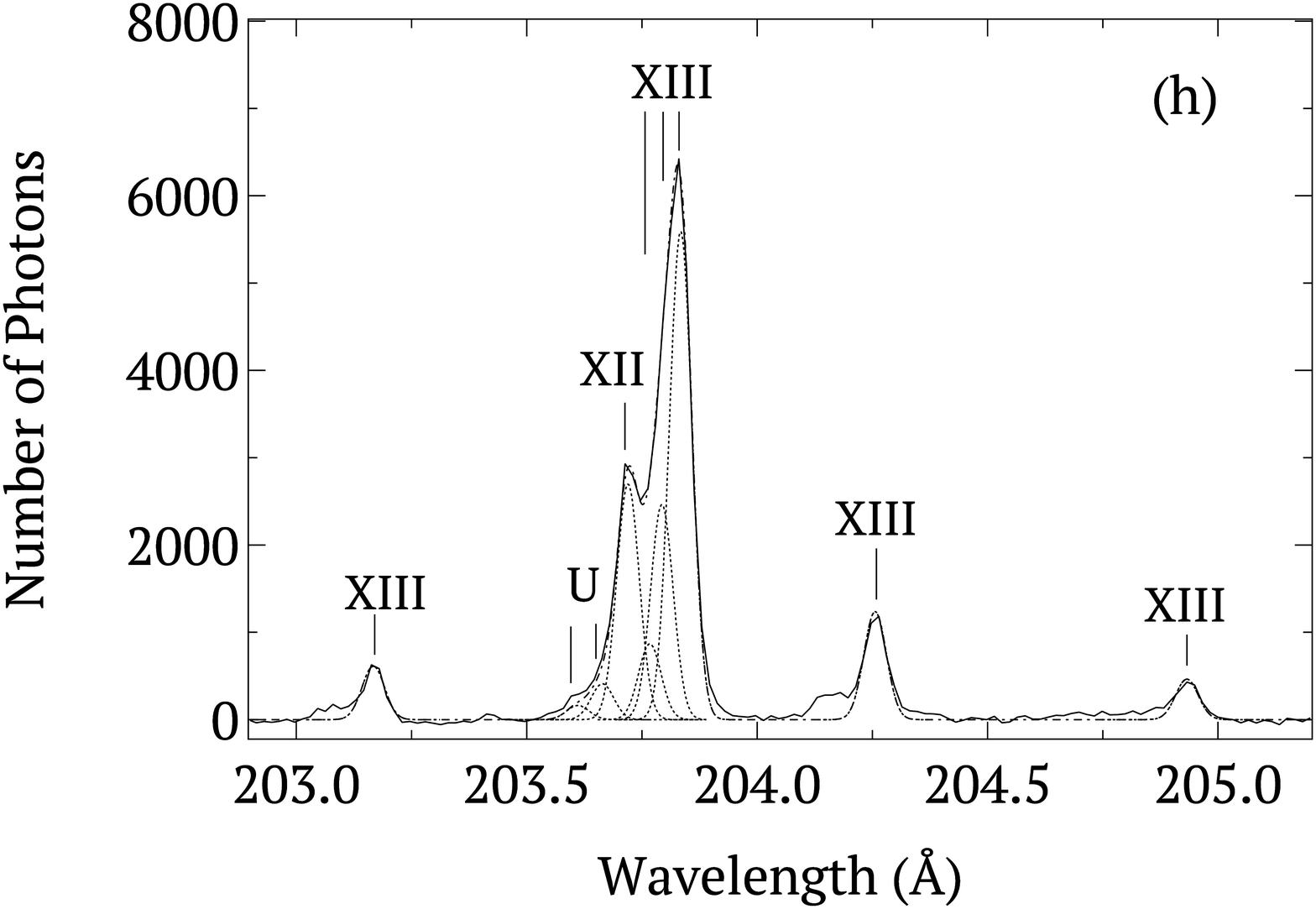}
        }%
    \end{center}
    \caption{\label{fig:fits1}
        Gaussian fits for spectral lines of interest from the 185--205~\AA\ range at $E_{\mathrm{e}} = 395$~eV and $I_{\mathrm{e}}=7$~mA. The measured spectrum is shown by the solid black curves. The Gaussian fits for individual lines are indicated by the gray curves. The sum of the individual fitted lines is plotted by the dot-dashed curve and is barely discernible from the measurements. Fe lines are indicated only by their spectroscopic Roman numeral. Identified oxygen and carbon lines labeled and unidentified lines are denoted by ``U''. 
     }%
   \label{fig:subfigures1}
\end{figure}

\begin{figure}[ht!]
     \begin{center}
        \subfigure{%
            \label{fig:nine}
            \includegraphics[width=0.4\textwidth]{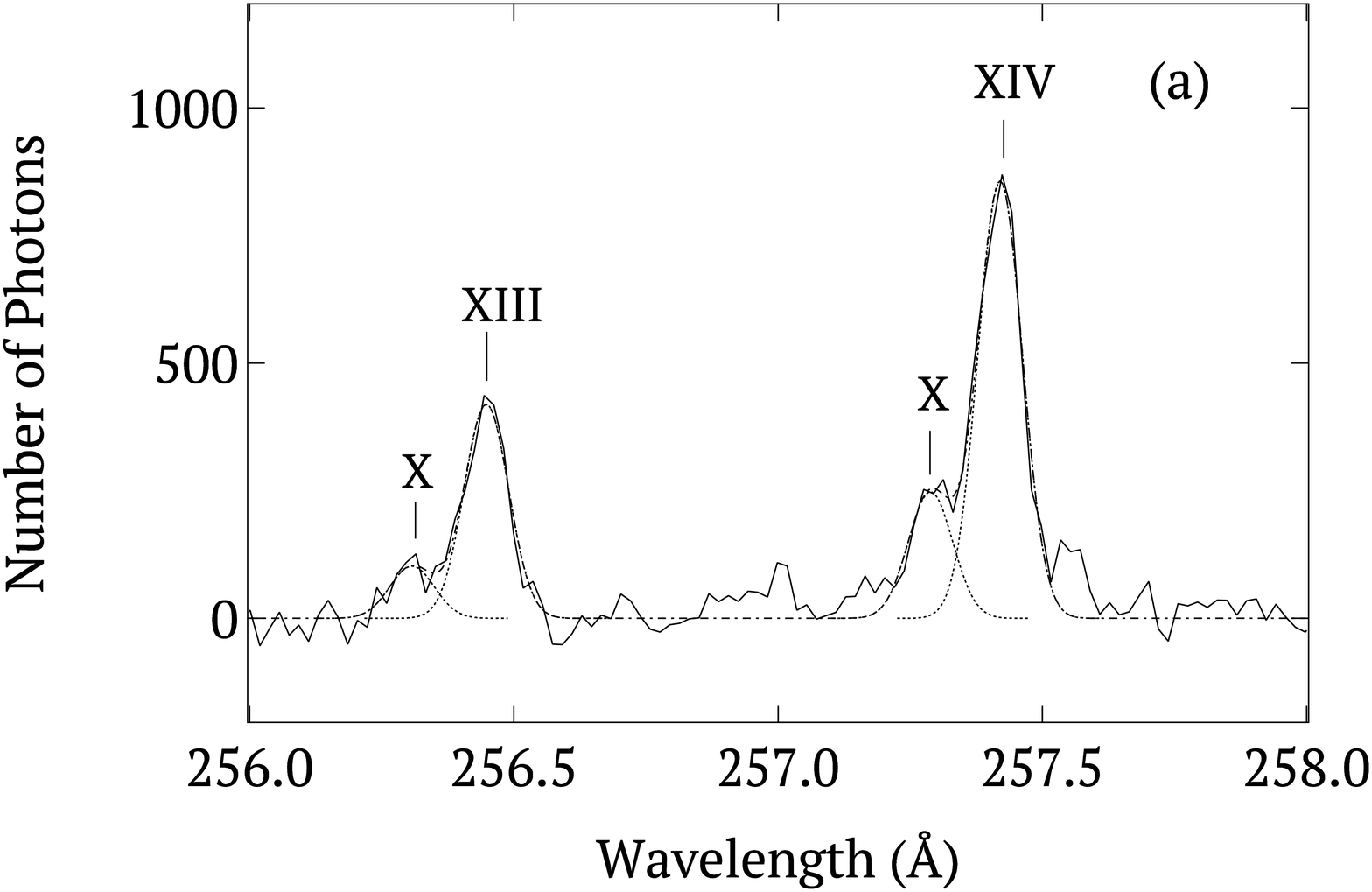}
            
        }%
        \subfigure{%
           \label{fig:ten}
           \includegraphics[width=0.4\textwidth]{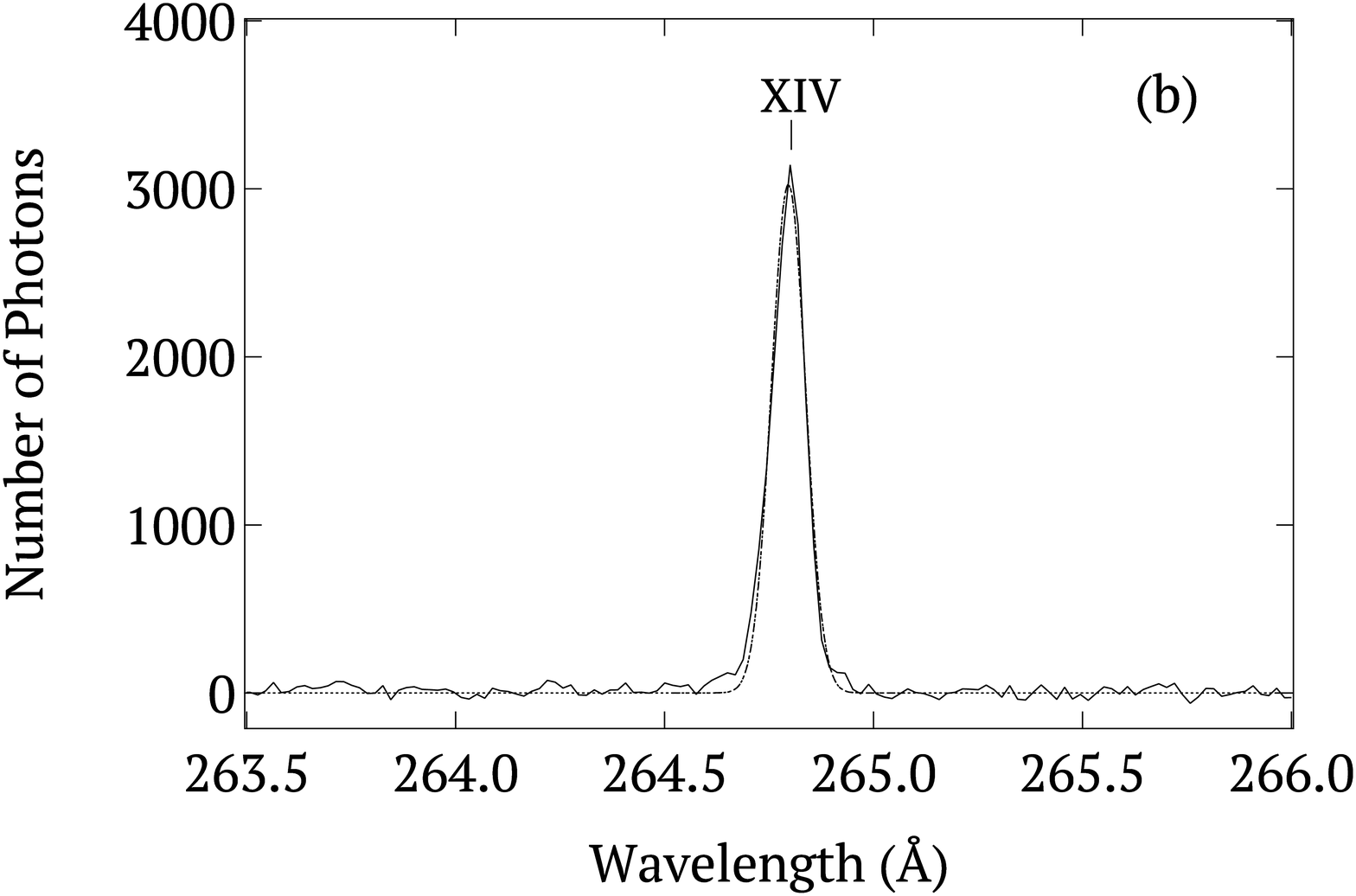}
       }\\ 
        \subfigure{%
            \label{fig:eleven}
            \includegraphics[width=0.4\textwidth]{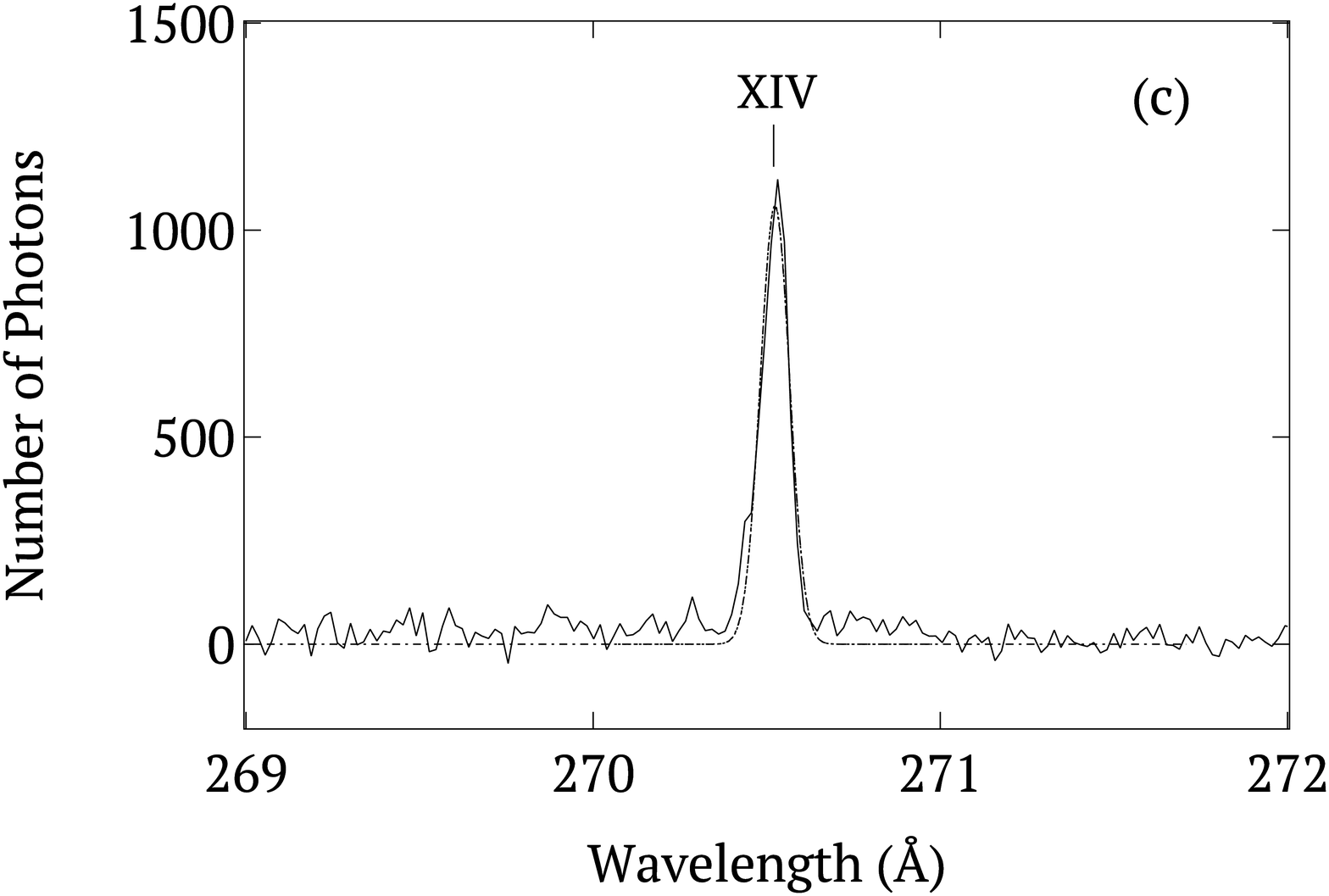}
        }%
        \subfigure{%
            \label{fig:twelve}
            \includegraphics[width=0.4\textwidth]{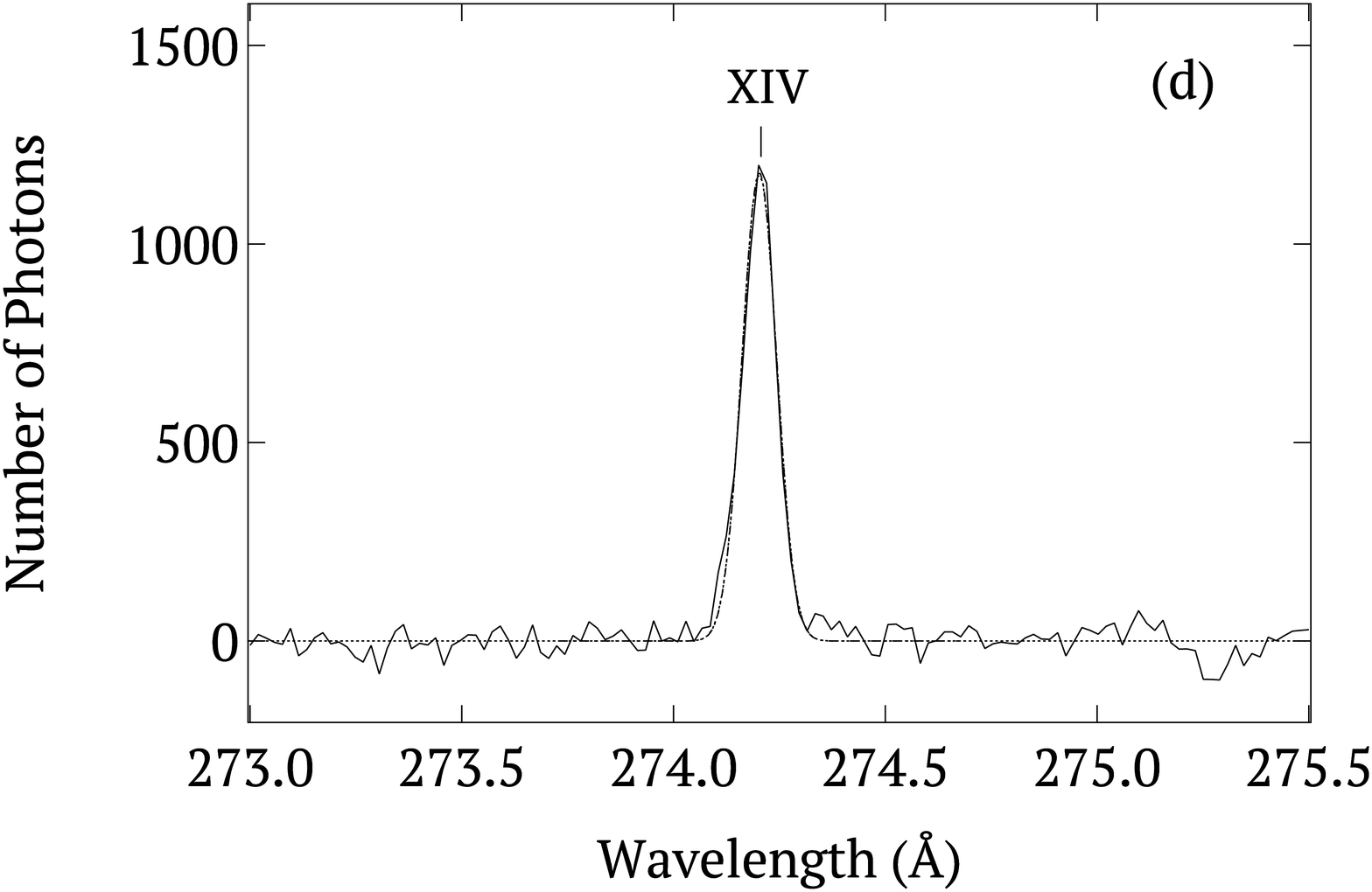}
        }%

    \end{center}
    \caption{\label{fig:fits2}
        Same as Figure~\ref{fig:fits1} for the 255--276~\AA\ range. 
     }%
   \label{fig:subfigures2}
\end{figure}

Figures~\ref{fig:fits1} and \ref{fig:fits2} illustrate example fits for $E_{\mathrm{e}}=395$~eV and $I_{\mathrm{e}}=7$~mA in the 185--205 and 255--276~\AA\ wavelength ranges, respectively. The corresponding extracted spectral line intensities are listed in Tables~\ref{table:table3}, \ref{table:table4}, and \ref{table:table5}. 

Intensities measured in EBIT can sometimes differ from theoretical predictions or astrophysical observations due to polarization effects. These effects arise because the ions in EBIT are excited by a directed electron beam. The resulting emission may not be isotropic, but we measure it only from a direction perpendicular to the electron beam axis. In contrast, theoretical calculations model the total intensity over all $4\pi$ solid angle. Moreover, most natural systems do not have such a preferred axis as exists in the experiment. The effect of this anisotropic emission is quantified by the polarization $P$ given by 
\begin{equation}
P=\frac{I_{\parallel} - I_{\perp}}{I_{\parallel}+I_{\perp}}, 
\label{eq:poldef}
\end{equation}
where $I_{\parallel}$ and $I_{\perp}$ represent the line intensities emitted parallel and perpendicular to the electron-beam axis. In the experiment, we observe at approximately 90$^{\circ}$ to the axis of the beam and denote the measured intensity as $I_{90}$. This measured intensity is related to the total intensity $I=I_{\parallel}+I_{\perp}$ emitted over all $4\pi$ solid angles by 
\begin{equation}
I_{90} \approx \frac{3I}{3-P}. 
\label{eq:polcorrect}
\end{equation}
\citet{Liang:ApJ:2009a} have calculated the polarization for many of the lines we have measured. For most of these lines, the polarization is only a few percent and the effects on the intensity are negligible and can be ignored. In addition, one needs to consider that the electrons in an EBIT spiral around the magnetic field lines with a pitch angle, $\theta_0$, measured relative to the field lines, that is given by $\epsilon=\sin^2\theta_0 = E_{\perp}/E_{\mathrm{e}}$ \citep{Gu:JPhysB:1999}. The resulting reduction in the polarization for electric dipole transitions is 
\begin{equation}
P^{\prime} = P \frac{2-3\epsilon}{2-\epsilon P}, 
\label{eq:pprime}
\end{equation}
in EBIT-I, $E_{\perp} \sim 50$--200~eV \citep{Levine:NIMB:1989, Beiersdorfer:PRA:1992, Gu:JPhysB:1999, Beiersdorfer:PRE:2001}. The largest predicted polarization here is for the Fe~\textsc{xiii} 202.04~\AA\ line with $P=21\%$. Taking the spiraling of the electrons into account reduces this to 6--18\% for the beam energies considered here. This is discussed in more detail in Section~\ref{subsec:FeXIII}. 


\subsection{Visible Light Imaging of the Ion Cloud}\label{subsec:ioncloud}

Visible lines from metastable ions were used to determine the size of the trapped ion cloud. Because visible light arises from levels that have long lifetimes of $\sim 10^{-3}$~s, the metastable ion emission spans the entire ion cloud and gives an accurate measurement of its dimensions. 
Since we are concerned with density-diagnostic measurements for Fe~\textsc{xii}, \textsc{xiii}, and \textsc{xiv}, it would have been preferable to measure the individual ion cloud sizes for each charge state. However, the visible lines from metastable Fe~\textsc{xii} and \textsc{xiii} were too weak to be detected. We were only able to measure the metastable Fe~\textsc{xiv} $3s^2\,3p\:^{2}P_{3/2}- 3s^{2}\,3p\:^{2}P_{1/2}$ transition at 5302.9\AA. This emission was isolated by using a 30~\AA\ bandpass filter centered at 5320~\AA. 
The lifetime of the Fe~\textsc{xiv} 5302.9~\AA\ transition is 16.7 $\mathrm{ms}$ \citep{Beiersdorfer:ApJ:2003, Brenner:PRA:2007}, which is long enough that the emission spans the ion cloud. 

The measurement of the Fe~\textsc{xiv} ion cloud diameter is expected to provide a reasonable approximation for nearby charge states. This can be understood in terms of the electrostatic and magnetic trapping that constrains the ion orbits. For electrostatic trapping, the maximum extent of the ion orbits is set by the balance between the ion kinetic energy and the ion charge $q$ multiplied by the electrostatic potential $\phi$. The ion temperatures in EBIT-I have been estimated to be $\approx 10q$~$\mathrm{eV}$ \citep{Beiersdorfer:RSI:1995}. However, the potential energy depth of the trap, $q \phi$, is also proportional to $q$. Consequently, for electrostatic trapping of a given element the ion cloud radius should be independent of charge since the potential and kinetic energies of the ions are both proportional to $q$ and so the radial turning points of their orbits are all the same. There is also magnetic trapping, in which the ion orbit radius is roughly given by the gyroradius, $mv/qB$, where $m$ is the ion mass, $v$ is the ion velocity, and $B$ is the magnetic field strength. For trapped ions the potential energy balances the kinetic energy so that the kinetic energy is $\propto q \phi$, which implies that $v \propto \sqrt{q}$ from which it follows that the gyroradius is $\propto 1/\sqrt{q}$. Thus, for magnetic trapping, the ratio of the ion orbits from Fe~\textsc{xii} and \textsc{xiv} is $\sqrt{13/11} \approx 1.09$ and we expect less than a $10\%$ difference between the Fe~\textsc{xii}, \textsc{xiii}, and \textsc{xiv} ion cloud radii. 

The apparatus and procedure for measuring the ion cloud have been previously reported by \citet{Arthanayaka:RSI:2018}, however the earlier setup suffered from some optical aberrations and so we have made several improvements. To enhance the image quality, we have replaced the 2 inch diameter convex lens with a high optical quality 4 inch diameter lens. Furthermore, a 1 inch diameter aperture was placed directly in front of the lens to minimize spherical aberration. Additional precautions were taken to minimize internal reflections of the optical system. The bandpass filter was placed between the EBIT-I window and the convex lens. For our setup, the ion cloud was measured with a magnification of 3.05  $\pm $ 0.06. 

\begin{figure}
	\centering \includegraphics[width=0.9\textwidth]{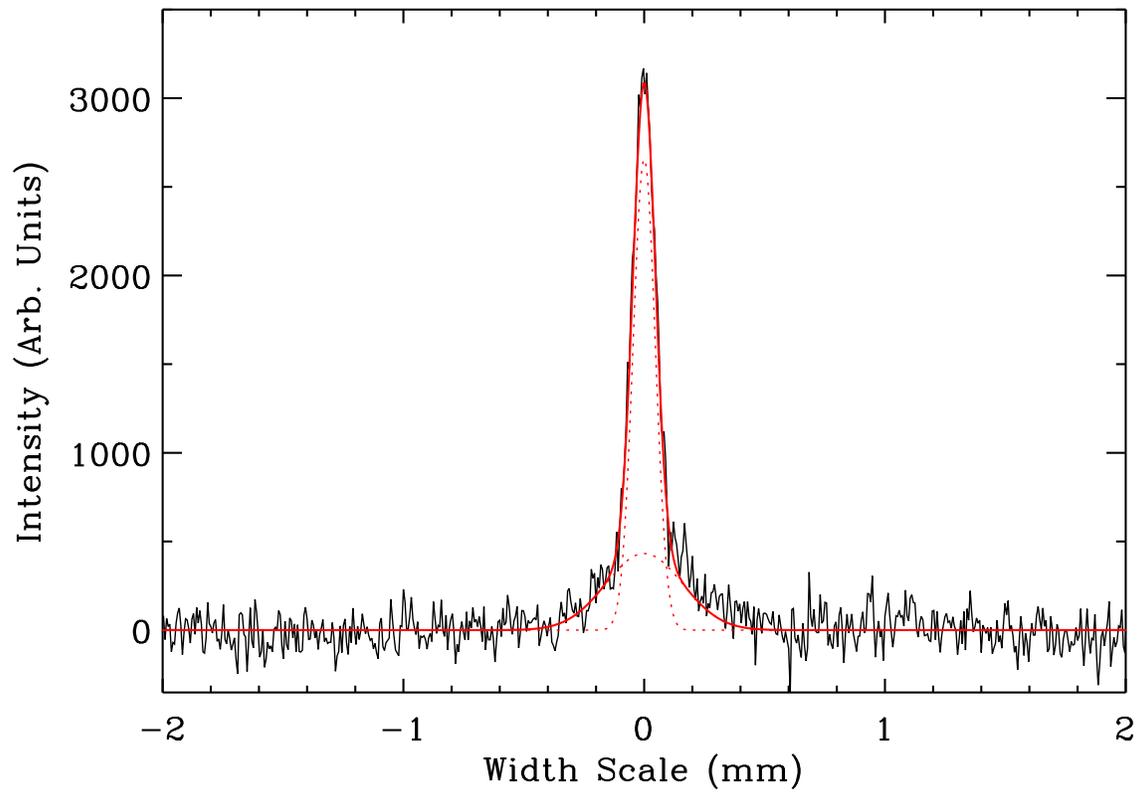}
	\caption{\label{fig:optical1} Lineout of the ion cloud image is shown by the solid black curve. These data were obtained for $E_{\mathrm{e}}$ = 395 eV and $I_{\mathrm{e}}$ = 7 mA.  The solid red curve illustrates the fit to the data, which is the sum of the two Gaussian components shown by the red dotted curves.
}
\end{figure}

Figure~\ref{fig:optical1} shows a line-out of an ion cloud measurement, which was obtained for $E_{\mathrm{e}} = 395$~eV and $I_\mathrm{e} = 7$~mA. 
In order to calculate the FWHM of the ion cloud, $\Gamma_{\mathrm{i}}$, the horizontal scale in pixels was converted to physical units based on the optical setup magnification and the pixel spacing of the CCD \citep[see][]{Arthanayaka:RSI:2018}. The ion cloud profile is dominated by a narrow Gaussian with a FWHM measured to be $\Gamma_{1} \approx 110$~$\mu\mathrm{m}$ (Tables~\ref{table:table6} and \ref{table:table7}). The second Gaussian component has a much smaller amplitude that is only $\approx 10$\% of the amplitude of the narrow component, but it is much broader with a FWHM of $\Gamma_{2}\approx 400$~$\mu\mathrm{m}$. There is some question as to whether the low amplitude broad component is real or a systematic error due to residual optical aberrations. Our experience has shown that improvements to the optical system have substantially reduced the amplitude of the broad component, but it was not possible to eliminate the broad component entirely. We have attempted to model the ion cloud by calculating ion trajectories in EBIT-I. As discussed in Section~\ref{subsec:orbit}, our model for the ion cloud qualitatively reproduces both the narrow and broad component structure of the observed ion cloud. For these reasons, we believe that the broad component is real and contains about 30\% of the ions. We found no systematic variations versus beam energy or beam current for either Gaussian component. 



\subsection{Electron Density} \label{subsec:density}

The electron beam is nearly Gaussian, with the density varying radially as
\begin{equation}
n_{\mathrm{e}}(r) = n_{\mathrm{0}}e^{-r^2/2\sigma_{\mathrm{e}}^2}, 
\label{eq:negauss}
\end{equation}
where $n_{0}$ is the central density of the beam and $\sigma_{\mathrm{e}}$ is the Gaussian beam width. It is useful to characterize the electron beam density using an average value, which we take to be the density that the beam would have if it were a uniform cylindrical electron beam of radius $r_{\mathrm{e}}$. We refer to this as the geometric average denoted by the symbol $\bar{n}_{\mathrm{e}}$. 

For a uniform cylindrical electron beam, $\bar{n}_{\mathrm{e}}$ is related to the electron current, $I_{\mathrm{e}}$, by 
\begin{equation}
\bar{n}_{\mathrm{e}} = \frac{I_{\mathrm{e}}}{\pi r_{\mathrm{e}}^2 e v_{\mathrm{e}}},
\label{eq:nebar}
\end{equation}
where $r_{\mathrm{e}}$ is the radius of the beam and $v_{\mathrm{e}}$ is the axial velocity of the electrons. Experimentally, $I_{\mathrm{e}}$ is measured by an ammeter connected to the EBIT collector and $v_{\mathrm{e}}$ is calculated from the space-charge-corrected electron beam energy $E_{\mathrm{e}}$. 

There remains a choice as to how we should define $r_{\mathrm{e}}$ for the electron beam and the corresponding equivalent cylindrical beam from which we derive our average. Here, we choose to define $r_{\mathrm{e}} = 2\sigma_{\mathrm{e}}$, so that 95\% of the electrons in the beam are within this radius. Alternative conventions are also possible, for example \citet{Liang:ApJ:2009} defined their beam radius to be the radius that enclosed 80\% of the beam electrons. Our choice of $r_{\mathrm{e}} = 2\sigma_{\mathrm{e}}$ has the property that the central density $n_{0}$ is twice the geometric average density $\bar{n}_{\mathrm{e}}$. Moreover, the density of the Gaussian beam at $r_{\mathrm{e}}$ is approximately one fourth of $\bar{n}_{\mathrm{e}}$. Thus, we can characterize our beam density by $\bar{n}_{\mathrm{e}}$.

The measurements of the geometry of the electron beam have been described in \citet{Arthanayaka:RSI:2018}. Briefly, we use the measured spectroscopic EUV line shapes to infer the electron beam profile. This works because the EUV emission comes from allowed transitions that are excited and decay within the electron beam. The typical lifetimes of the transitions are $10^{-10}$~s and the thermal energies of the ions in EBIT are estimated to be $\sim 10q \sim 10^{2}$~eV, which corresponds to a velocity of $3 \times 10^{4}$~$\mathrm{m\,s^{-1}}$. For a typical 60~$\mu$m electron beam width, the crossing time of an iron ion is about $10^{-9}$~s, which is much shorter than the lifetimes of the EUV-radiating levels. Doppler broadening of the spectral lines is small and adds in quadrature with the broadening due to the finite width of the emission source. As a result, the widths of the EUV spectral lines reflects mainly the width of the electron beam. 
As these contributions add in quadrature, Doppler broadening contributes up to 10\% of the line widths, which we incorporate into the analysis as a $10\%$ uncertainty in the electron beam width. 
Gaussian profiles were fit to the EUV spectral lines in order to infer the electron beam FWHM $\Gamma_{\mathrm{e}} = 2.355\sigma_{\mathrm{e}}$. For these fits, the $x$-axis of the spectra was calibrated in physical units via the pixel size, rather than in wavelength units. To determine the beam size, strong lines were used that were less affected by blends as described in \citet{Arthanayaka:RSI:2018}. The beam width was derived for each beam current and energy studied. There was some slight variation as a function of current and energy, with values in the range of $\Gamma_{\mathrm{e}} \approx 50$--$60$~$\mu$m (Tables~\ref{table:table6} and \ref{table:table7}). This is consistent with previous measurements on EBIT-I \citep{Marrs:PhysScr:1995, Utter:NIMA:1999}. 



 
\section{Effective Electron Density}\label{sec:effdensity}

The trapped ions pass through the electron beam for only a small portion of their trajectories. As a result, the trajectories of the trapped ions produce an ion cloud that is larger in radius than that of the electron beam. For this reason, Previous studies have suggested that the relevant density for line ratios in an EBIT experiment is the spatially averaged density that the ions experience along their orbits, the so-called effective density $n_{\mathrm{eff}}$ \citep{Crespo:CanJPhys:2002, Liang:ApJ:2009a, Nakamura:ApJ:2011, Shimizu:AA:2017}. If the ion cloud is Gaussian with the same centroid as the electron beam, then the electron density spatially averaged over the ion cloud is given by 
\begin{equation}
n_{\mathrm{eff}} = \frac{4 \ln{(2)} I_{\mathrm{e}}}{\pi e v_{\mathrm{e}}} \left(\frac{1}{\Gamma_{\mathrm{e}}^2 + \Gamma_{\mathrm{i}}^2}\right). 
\label{eq:neff}
\end{equation}
This expression is derived in Appendix~\ref{app:appendix1}. 
For our experiment, the ion cloud has a double-Gaussian structure, which leads to a more complicated expression: 
\begin{equation}
n_{\mathrm{eff}} = \frac{4 \ln{(2)} I_{\mathrm{e}}}{\pi e v_{\mathrm{e}}}
\left(\frac{1}{A_1\Gamma_{1} + A_2\Gamma_{2}}\right)
\left( \frac{A_1 \Gamma_1}{\Gamma_{1}^2 + \Gamma_{\mathrm{e}}^2} + \frac{A_2 \Gamma_{2}}{\Gamma_{2}^2 + \Gamma_{\mathrm{e}}^2}\right), 
\label{eq:neffdbl}
\end{equation}
which is also derived in Appendix~\ref{app:appendix1}. Here, $A_1$ and $A_2$ are the amplitudes of the measured ion cloud components and $\Gamma_1$ and $\Gamma_2$ are the corresponding FWHMs. Note that in Equation~(\ref{eq:neffdbl}) these are the observed amplitudes projected onto the detector, which differ by a multiplicative scaling factor from the amplitudes for the ion distribution as a function of radius. 

In reality the relevant quantity is the time-averaged density experienced by the collection of ions, rather than the spatially averaged density described above. This is because the process of excitation and relaxation that determines the level populations depends on density as a function of time. In general, the average given in Equations~(\ref{eq:neff}) and (\ref{eq:neffdbl}) are not the same as the time-averaged density experienced by any particular ion along its trajectory. However, as long as the ions are formed at random initial locations in the electron beam, then their trajectories are uncorrelated and the spatial averages given by Equations~(\ref{eq:neff}) and (\ref{eq:neffdbl}) are equivalent to the time average of each ion averaged over the collection of ions in the trap. This can be deduced analytically as is shown in Appendix~\ref{app:appendix2}. We have also verified that it is true using the ion-trajectory calculations that are described in Section~\ref{subsec:orbit}. 


We measured the two components making up $\Gamma_{\mathrm{i}}$ and the distribution of the ion cloud using emission from Fe~\textsc{xiv}, as discussed above in Section~\ref{subsec:ioncloud} and in \citet{Arthanayaka:RSI:2018}. As mentioned above, the dominant component of the ion cloud FWHM was typically $\Gamma_{\mathrm{i}} \approx 110$~$\mu\mathrm{m}$. For a typical electron beam radius of $\Gamma_{\mathrm{e}} \approx 55$~$\mu \mathrm{m}$, the effective density is about 30\% of the nominal beam density, $n_{\mathrm{eff}}\approx 0.3 \bar{n}_{\mathrm{e}}$. Tables~\ref{table:table6} and \ref{table:table7} give the measured $\Gamma_{1}$ and $\Gamma_{2}$ values and the inferred $n_{\mathrm{eff}}$ corresponding to each specific experimental set of conditions. 

When we plotted the density sensitive line ratios as a function of $n_{\mathrm{eff}}$ we found several discrepancies between the experiment and the calculations. These are discussed in more detail below, in Section~\ref{sec:results}. Some of those discrepancies would disappear if we were to plot the line ratios versus the nominal beam density, $\bar{n}_{\mathrm{e}}$, instead. This prompted us to consider whether or not $n_{\mathrm{eff}}$ really is the correct density to compare with. A two-level analytic model demonstrates that it is, in fact, correct to use the effective density. 

\subsection{Two-Level Analytic Model}\label{subsec:twolevel} 

For a two-level model, let $f$ be the relative population of the upper level so that $1-f$ is the population of the lower level. Also, $n_{\mathrm{e}}$ is the electron density, $C$ is the collisional excitation rate coefficient, $D$ is the collisional de-excitation rate coefficient, and $A_r$ is the radiative de-excitation rate. Then the differential equation describing the level population is, 
\begin{equation}
\frac{df}{dt} = n_{\mathrm{e}}C(1-f) - n_{\mathrm{e}}Df - A_r f.
\label{eq:ode1}
\end{equation}
This equation can be solved analytically. The equilibrium level is given by $df/dt=0$ so that 
\begin{equation}
f_{\mathrm{eq}}= \frac{n_{\mathrm{e}}C}{n_{\mathrm{e}}(C+D)+A_r}. 
\label{eq:hieq}
\end{equation}
Defining an initial condition that $f(t=0)=f_{0}$, the solution to Equation~(\ref{eq:ode1}) is: 
\begin{equation}
f(t) = \frac{n_{\mathrm{e}}C}{n_{\mathrm{e}}(C+D)+A_r}\left(1-e^{-[n_{\mathrm{e}}(C+D)+A_r]t}\right) + f_{0}e^{-[n_{\mathrm{e}}(C+D)+A_r]t}.
\label{eq:hisol}
\end{equation}


For the low-density case, collisions are negligible so that
\begin{equation}
\frac{df}{dt} = -A_r f. 
\label{eq:odelo}
\end{equation}
The equilibrium solution for the low density case is $f = 0$, indicating that all the ions are in the ground state. As a function of time, the evolution to this ``coronal limit'' is given by  
\begin{equation}
f(t) = f_{0}e^{-A_r t}. 
\label{eq:losol}
\end{equation}

Now, consider an ion passing in and out of the beam and starting in the low-density equilibrium with $f(0)=0$. Suppose the ion passes into the beam for a time $\Delta t_{\mathrm{hi}}$ and then is outside the beam for a time $\Delta t_{\mathrm{lo}}$ with the sequence repeating many times. Let the index $i$ refer to the time-step for completing the entire sequence so that $t_i = i(\Delta t_{\mathrm{lo}} + \Delta t_{hi})$. It is also convenient to introduce a short-hand for the exponentials, so we set $\eta = e^{-[n_{\mathrm{e}}(C+D)+A_r]\Delta t_{\mathrm{hi}}}$ and $\lambda = e^{-A_r\Delta t_{\mathrm{lo}}}$. At each time-step, the level populations are given by
\begin{equation}
f_{i}=\frac{n_{\mathrm{e}}C}{n_{\mathrm{e}}(C+D)+A_r}(1-\eta) \lambda + \eta \lambda f_{i-1}.
\label{eq:popiter}
\end{equation}
In equilibrium it must be that $f_{i} = f_{i-1} = f_{\infty}$. Substituting into Equation~(\ref{eq:popiter}) and solving for $f_{\infty}$ we find that 
\begin{equation}
f_{\infty} = \frac{n_{\mathrm{e}}C}{n_{\mathrm{e}}(C+D)+A_r} \frac{(1-\eta)\lambda}{1-\eta\lambda}. 
\label{eq:sweq}
\end{equation}
Since the population change at each $\Delta t_{\mathrm{hi,lo}}$ is very small, we can use a Taylor series expansion for the exponential terms so that $\eta \approx 1-[n(C+D)+A]\Delta t_{\mathrm{hi}}$ and $\lambda \approx 1 - A\Delta t_{\mathrm{lo}}$. Substituting these into Equation~(\ref{eq:sweq}) and eliminating second order terms gives
\begin{equation}
f_{\infty} = \frac{n_{\mathrm{e}}C}{n_{\mathrm{e}}(C+D)+A} 
\frac{[n_{\mathrm{e}}(C+D)+A]\Delta t_{\mathrm{hi}}}{ [n_{\mathrm{e}}(C+D)+A]\Delta t_{\mathrm{hi}} + A \Delta t_{\mathrm{lo}}}. 
\label{eq:sweq1}
\end{equation}
This is the high-density equilibrium population $n_{\mathrm{e}}C/[n_{\mathrm{e}}(C+D)+A_r]$ multiplied by a correction factor. 


The effective density is defined as the time-averaged density experienced by the ion. In this example, the ion travels from the high-density electron beam with density $n_{\mathrm{e}}$ for time $\Delta t_{\mathrm{hi}}$ and then is outside the beam where the density is $0$ for a time $\Delta t_{\mathrm{lo}}$. So we have, 
\begin{equation}
n_{\mathrm{eff}} = \frac{n_{\mathrm{e}} \Delta t_{\mathrm{hi}}}{\Delta t_{\mathrm{hi}} + \Delta t_{\mathrm{lo}}}. 
\label{eq:neffdef}\
\end{equation}
Re-writing Equation~(\ref{eq:sweq1}) in terms of $n_{\mathrm{eff}}$ gives
\begin{equation}
f_{\infty} = \frac{n_{\mathrm{eff}}C}{n_{\mathrm{eff}}(C+D)+A}. 
\label{eq:sweq2}
\end{equation}
But, this is the equilibrium level population if the ions are always in a region where the electron density is $n_{\mathrm{eff}}$. This demonstrates that the effective density controls the level populations. 

\subsection{Ion Orbit Calculations}\label{subsec:orbit}

We have calculated the ion orbit trajectories in order to determine the amount of time that ions spend inside and outside the beam and to better understand the observed shape of the ion cloud. The trajectories are calculated in the two-dimensional plane perpendicular to the beam axis. This is justified by the axial symmetry of EBIT. The electron beam is modeled as a Gaussian distribution, as given by Equation~(\ref{eq:negauss}) with $n_{0}$ and $\sigma_{\mathrm{e}}$ set by the experimentally measured values. The electron beam produces an electric-field vector, which can be determined analytically at any location. The experiment also has an axial 3~T magnetic field, which is also included in the model. Ion trajectories were calculated using test particles by integrating the equation of motion for the Lorentz force from the combined axial magnetic field and radial electric field. As a result, ion space-charge effects are ignored, but these are expected to be small since the density of ions is low. The integration was carried out using the Boris algorithm \citep{Birdsall:book, Qin:PoP:2013}. The results were tested for energy conservation and found to conserve energy to about a few percent over at least 10~ion cyclotron periods. We compare to this timescale because it is roughly similar to the timescales for the radial oscillations of the ions. The method is very fast allowing for 1000 ion trajectories to be calculated in minutes.

Test particle Fe$^{13+}$ ions were simulated with experimentally motivated initial conditions. The initial velocities were drawn at random from a Maxwellian distribution with a temperature of $130$~eV. The initial positions were assumed to be proportional to the local electron density, as we expect the formation rate to be proportional to the electron beam density profile. For each simulated ion trajectory, we calculated the fraction of time that the ion spent at $r < r_{\mathrm{e}}=2\sigma_{\mathrm{e}}$, which we consider to be the time that the ion is ``in the beam''. We found that the ion trajectories typically spend about 40\% of their time in the electron beam. This is roughly consistent with our finding that the effective density is about 30\% of the average electron beam density. 

In order to simulate the ion cloud measurement, we chose a late time in the simulation so that the orbits had time to become incoherent and remove initial transients. At that time, we made a histogram of the number of ions that would be seen along a line of sight looking through the device. In practice this was done by making a histogram of the $x$-coordinate of the ions, rather than their radius $r$. This is the Abel transform of the radial distribution. We found that our simulated ion clouds could be described by two Gaussian components, a strong narrow component with $\Gamma \approx 60$~$\mu$m and a weaker broad component with $\Gamma \approx 300$~$\mu$m. The amplitude of the broad component is about 10\% that of the narrow component, and it contains about 30\% of the ions. These characteristics are very similar to the measured ion cloud. One difference with the experiment is that both components in the simulated distribution have slightly smaller widths than in the experiment, which suggests that some ions in the model are more strongly confined to the electron beam than they are in the experiment. 
Although we do not reproduce the precise widths of the observed ion clouds, it is interesting that the model reproduces the general structure with roughly similar values for the widths of both the narrow and broad components and similar amplitudes. These results support the existence of the broad component seen in the experiment. 

A possible explanation for the differences between the simulated versus measured ion clouds is that the ion-velocity distribution that we use to initialize the ions in the simulation could be inaccurate. In reality, the ions could have a greater temperature than we assumed or they could have a non-thermal distribution. Alternatively, there could be collisions in the electron beam that tend to scatter the ions to wider orbits. 
 
\section{Model Calculations} \label{sec:factheory}

In order to compare our experimental line ratio results with theory, we have calculated the predicted line ratios using the Flexible Atomic Code \citep[FAC;][]{Gu:CanJPhys:2008}. FAC is a widely used software package that performs fully relativistic atomic calculations. Calculations were performed for all three ions. The atomic model for these ions included all levels for principal quantum number $n \leq 5$ and included cascades among those levels. The electron-energy distribution was assumed to be a Gaussian centered at the beam energy of either 395~eV or 475~eV with a $1\sigma$ width of 50~eV \citep{Levine:NIMB:1989, Beiersdorfer:PRA:1992, Gu:ApJ:1999}. The level populations and model line intensities were then calculated over a range of densities based on the FAC-derived collisional and radiative transition rates. 

For solar physics applications, density diagnostics are often interpreted using the atomic data tabulated in CHIANTI. However, the CHIANTI atomic data are tabulated for a Maxwellian electron energy distribution, whereas in the experiment the electron beam is approximately monoenergetic. 
We found though, that the difference in the electron energy distributions used in the modeling make little difference and so the CHIANTI values and our FAC values agree to better than 10\% for the line-intensity ratios. This is likely because the collision calculations in both FAC and CHIANTI are in the high-energy Born-approximation limit. Density sensitivity is determined by the populations of the metastable levels, which have excitation energies on the order of several eV. For comparison with CHIANTI we used Maxwellian distributions that had $k_{\mathrm{B}}T_{\mathrm{e}}=430$~eV, where $k_{\mathrm{B}}$ is the Boltzmann constant and $T_{\mathrm{e}}=5$~MK is the electron temperature. Thus, in both CHIANTI and in the nearly monoenergetic distribution used in our FAC calculations, the average electron energy is much larger than the metastable level excitation energies. Consequently the atomic collision data for the metastable levels from both CHIANTI and FAC are similar. In the comparisons shown below (Section~\ref{sec:results}), our comparisons to the FAC calculations are the most appropriate. For reference, we also plot the ratios predicted by CHIANTI for a Maxwellian plasma at $T_{\mathrm{e}}=5$~MK. 

\section{Results} \label{sec:results}

\begin{figure}
	\centering \includegraphics[width=0.9\textwidth]{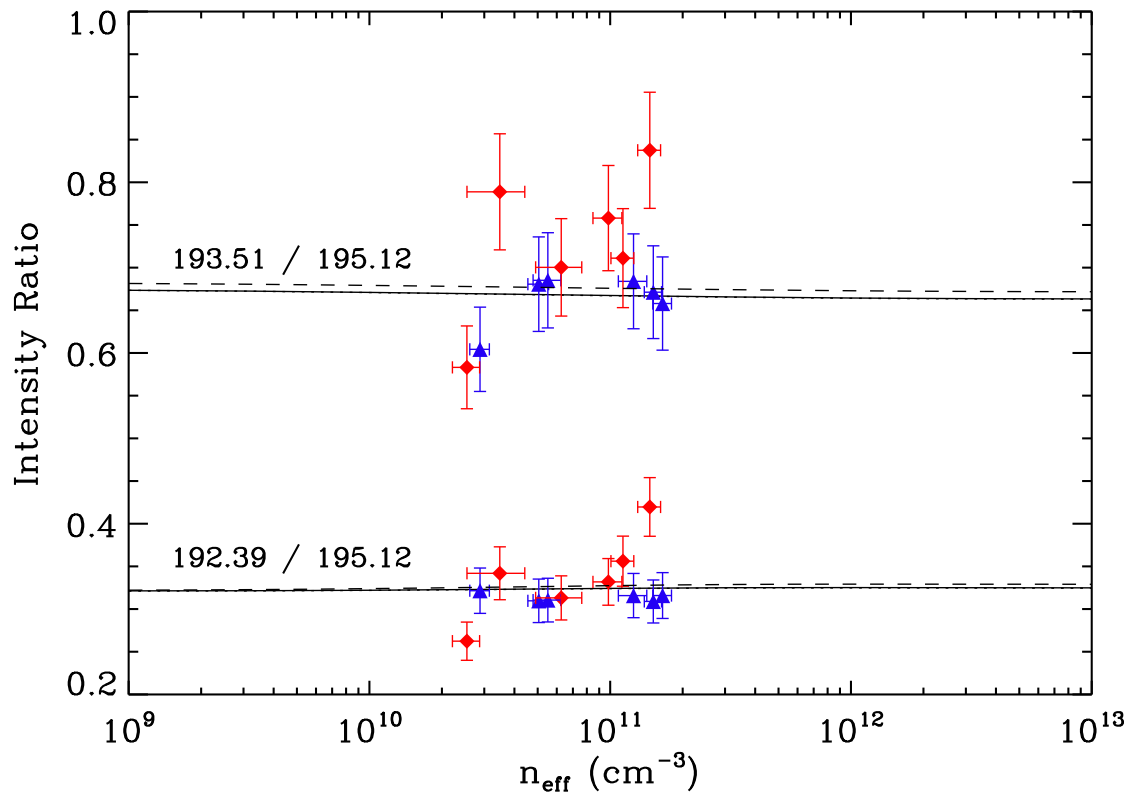}
	\caption{\label{fig:ratio1} Fe~\textsc{xii} density-independent line-intensity ratios. The data were obtained for $E_{\mathrm{e}}$ = 395 eV and  475 eV and are plotted using blue triangular and red tetragonal markers, respectively. The vertical error bars were estimated from the fitting using Gaussian line profiles and include the systematic uncertainty from the fitting. The horizontal error bars represent the uncertainty of the $E_{\mathrm{e}}$, $I_{\mathrm{e}}$, and $r_{\mathrm{e}}$ measurements. The dotted line and the solid lines represent the FAC calculations for $E_{\mathrm{e}}$ = 395 and 475 eV, respectively. These two lines are overlapping here. The dashed curve indicates the results from CHIANTI assuming a Maxwellian electron distribution at a temperature of 5~MK. 
}
\end{figure}

\begin{figure}
	\centering \includegraphics[width=0.9\textwidth]{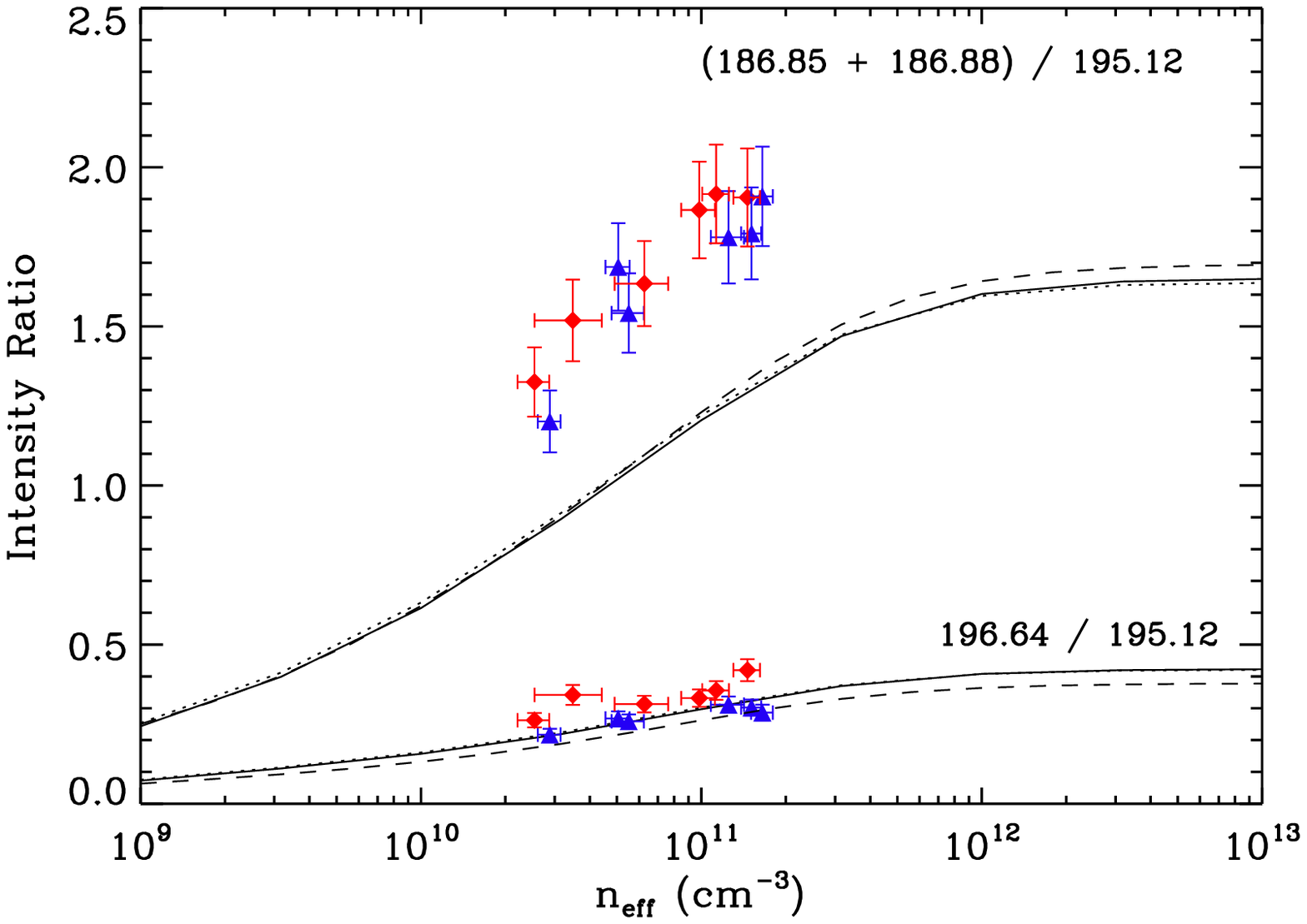}
	\caption{\label{fig:ratio2} Same as Figure~\ref{fig:ratio1} but for the Fe~\textsc{xii} density-dependent line-intensity ratios.
}
\end{figure}

\begin{figure}
	\centering \includegraphics[width=0.9\textwidth]{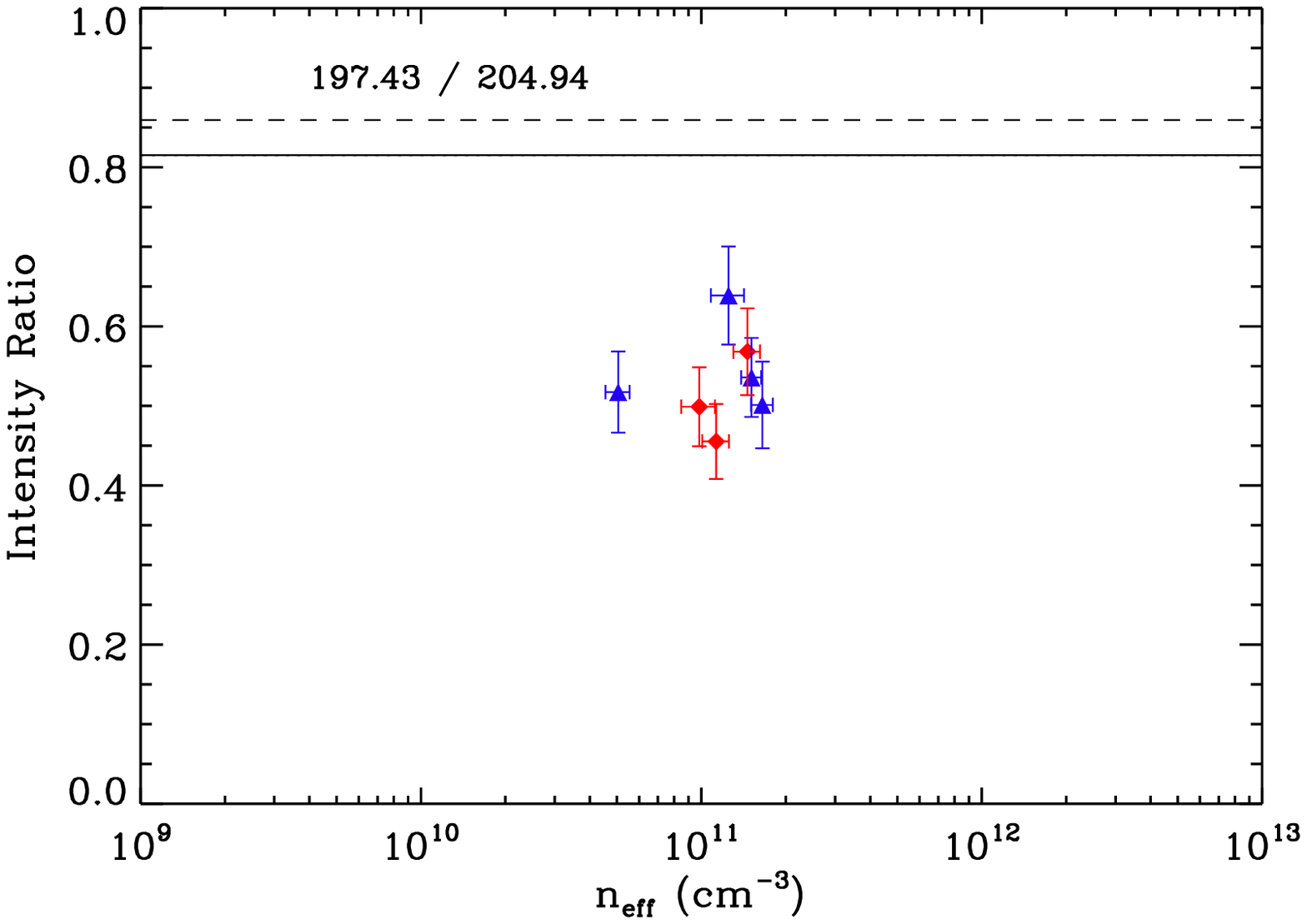}
	\caption{\label{fig:ratio3} Same as Figure~\ref{fig:ratio1}, but for the Fe~\textsc{xiii} density-independent line-intensity ratios. 
}
\end{figure}

\begin{figure}
	\centering \includegraphics[width=0.9\textwidth]{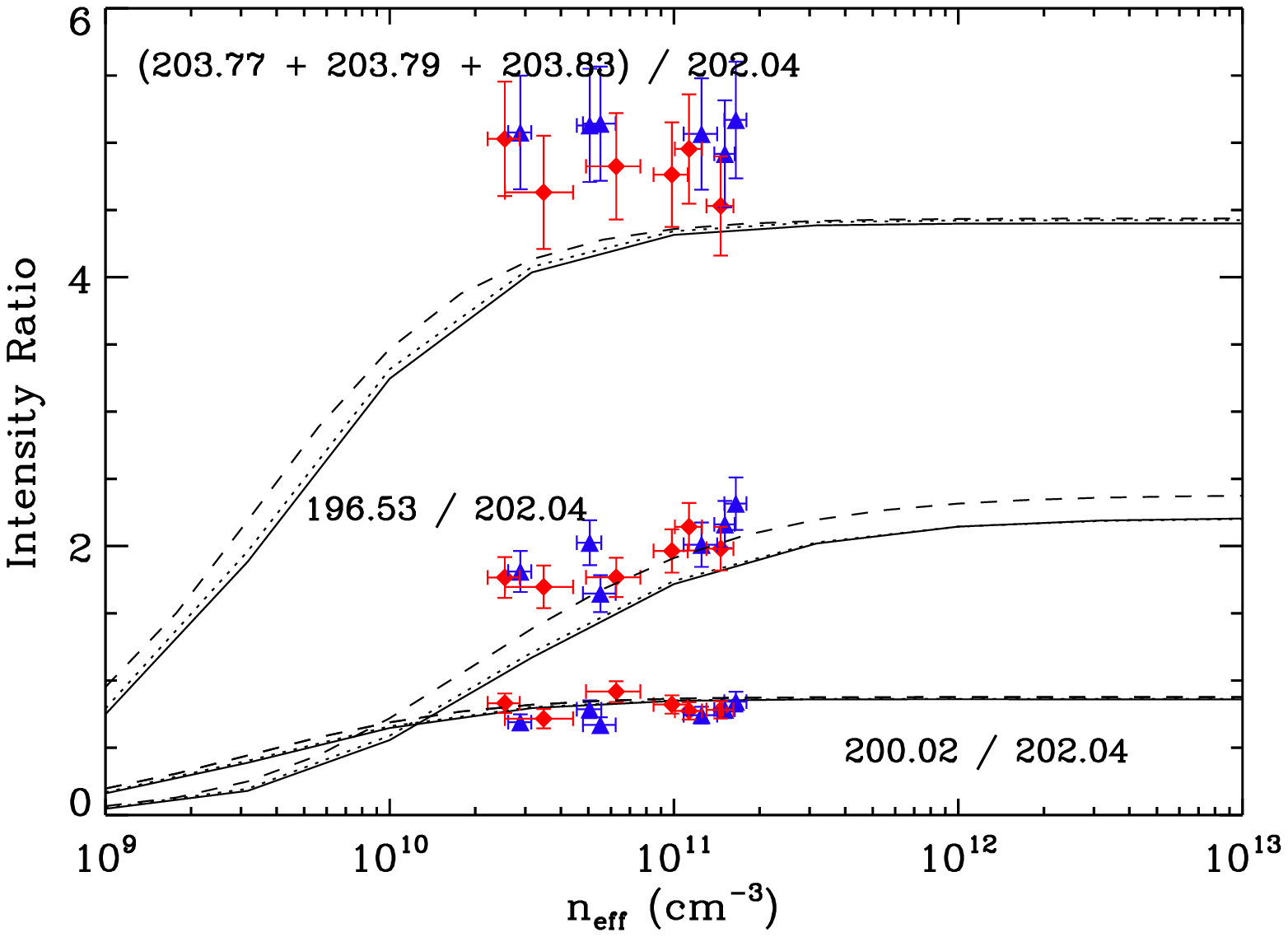}
	\caption{\label{fig:ratio4} Same as Figure~\ref{fig:ratio3}, but for the Fe~\textsc{xiii} density-dependent line-intensity ratios. 
}
\end{figure}

\begin{figure}
	\centering \includegraphics[width=0.9\textwidth]{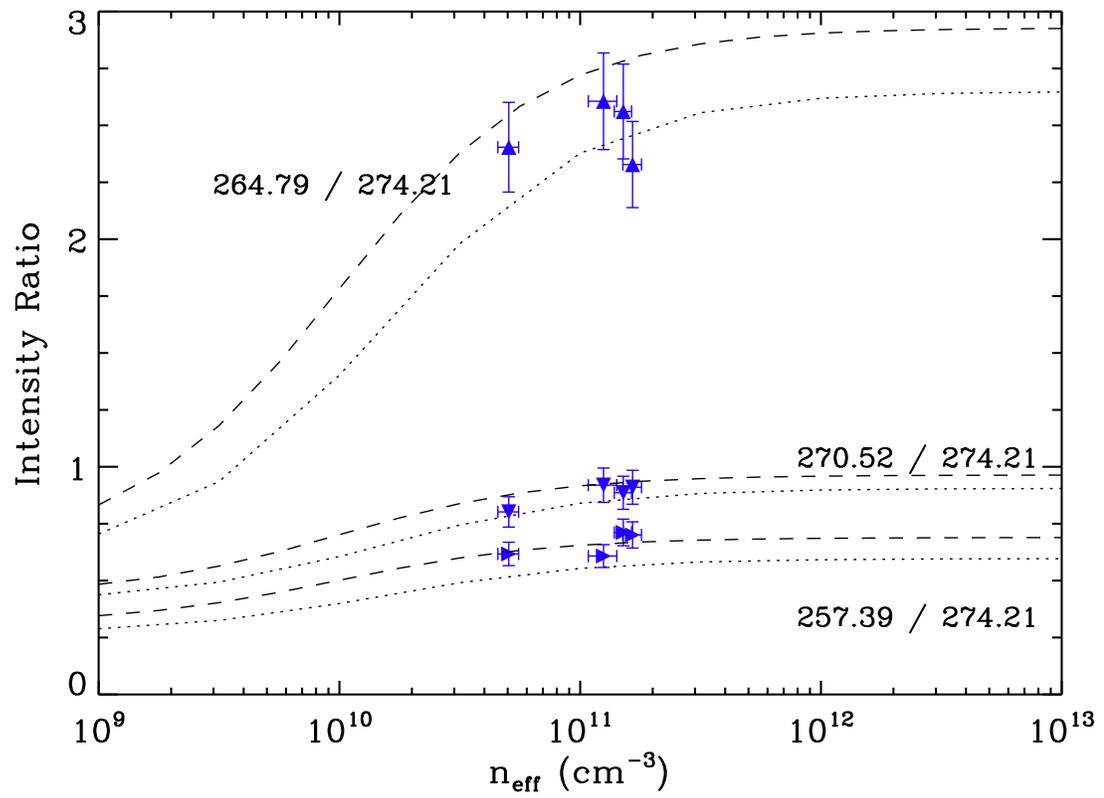}
	\caption{\label{fig:ratio5} Same as Figure~\ref{fig:ratio3}, but for the Fe~\textsc{xiv} density-dependent line-intensity ratios. The data were obtained for $E_{\mathrm{e}}=395$~eV only.
}
\end{figure}

The results for the Fe~\textsc{xii}--\textsc{xiv} line ratios are presented in Figures~\ref{fig:ratio1}--\ref{fig:ratio5}. The uncertainties of the experimental ratios are determined by propagating the error in the intensities derived from the Gaussian fits, described in Section~\ref{subsec:specline}. To draw the error bars, we have summed in quadrature the statistical uncertainties and the estimated 8\% systematic uncertainty that depends on the fitting model. The data are plotted as a function of the effective density $n_{\mathrm{eff}}$, as described above. There are uncertainties in $n_{\mathrm{eff}}$ from fitting a double-Gaussian shape to the ion cloud. The beam energy is also uncertain due to the space charge. We correct the energy for the space charge as described earlier and consider the uncertainty in $E_{\mathrm{e}}$ to be equal to the $\approx 20$~eV size of the correction. Propagating the fit parameter uncertainties and the space charge correction uncertainty into $n_{\mathrm{eff}}$ produces the horizontal error bars in Figures~\ref{fig:ratio1}--\ref{fig:ratio5}. 

\subsection{Fe~\textsc{xii}}\label{subsec:FeXII}

Fe~\textsc{xii} has several bright unblended lines that are density insensitive with respect to one another at 192.39, 193.51, and 195.12 \AA. For this reason, these lines are used to calibrate instruments and test theoretical models \citep{DelZanna:AA:2005}. We have used ratios among these lines for a similar purpose, to ensure the consistency of our results. Figure~\ref{fig:ratio1} demonstrates that there is no significant density sensitivity for ratios among these lines and that the predicted ratios are in good agreement with the calculations. The weighted value for our measurement of the 192.39/195.12 ratio is $0.68 \pm 0.02$ and the weighted mean for the 193.51/195.12 ratio was $0.32 \pm 0.01$. These may be compared to the theoretically predicted values of $0.67$ and $0.32$, respectively. Good agreement was also found for this line ration in the earlier EBIT measurements of \citet{Trabert:ApJS:2014}, though the density was not determined for those measurements.

Density diagnostics for Fe~\textsc{xii} can be formed by taking the ratio of a density sensitive line with any of the above three density-insensitive lines \citep{Flower:AA:1977, Dere:ApJS:1979, Schmitt:ApJ:1996, DelZanna:AA:2005, Young:PASJ:2007}. Two Fe~\textsc{xii} density sensitive lines were observed in the measured spectral range, the $3s^2\,3p^3 \:{^2}{D}{^o_{5/2}} - 3s^2\,3p^2\,(^2P)\,3d\:^{2}{F}_{7/2}$ transition at 186.88~\AA\ and the $3s^2\,3p^3\:^{2}{D}^{o}_{5/2}-3s^2\,3p^2\,(^{1}D)\,3d\:^{2}D_{5/2}$ transition at 196.64~\AA. The 186.88~\AA\ line is blended with the Fe~\textsc{xii} $3s^2\,3p^3\:^{2}{D}{^o_{3/2}} - 3s^2\,3p^2\,(^2P)\,3d\:^{2}F_{5/2}$ transition at 186.85~\AA. Because the blend is difficult to separate reliably, we study the ratio of the total intensity of both $(186.85+186.88)$~\AA\ lines relative to the 195.12~\AA\ line. The same procedure is commonly used in the analysis of solar observations. The Fe~\textsc{xii} line at 196.64~\AA\ is close to an Fe~\textsc{xiii} line at 196.53~\AA, but our spectral resolution is sufficient to clearly separate these lines, as can be seen in Figure~\ref{fig:fifth}. We note that the spectral resolution of EIS is similar to that of our experiment so that the 196~\AA\ blend is also separable in solar observations. 

Our results, for these two Fe~\textsc{xii} ratios are shown in Figure~\ref{fig:ratio2}. The intensity ratio for the 196.64/195.12 lines are in reasonable agreement with the predicted curve. For the (186.85 + 186.88)/195.12 ratio, the ratio versus density curve is significantly different from the calculation, with the measured ratio having a greater magnitude and a steeper slope versus density than predicted. The difference is above the estimated 20\% uncertainty for the calculations. The polarization for these lines is only a few percent \citep{Liang:ApJ:2009a} and polarization effects cannot account for the observed discrepancy with theory. 

When using the intensity ratio as a diagnostic, the inferred densities change exponentially as a function of the ratio, so the inferred density is very sensitive to small changes in the ratios. For the (186.85 + 186.88)/195.12 diagnostic, the interpretation of observed line intensities based on the current theoretical values would overestimate the density by about an order of magnitude.

\citet{Young:AA:2009} discussed these Fe~\textsc{xii} line ratios, $(186.85 + 186.88) /195.12$ and $196.64 /195.12$, in the solar spectrum and found that the derived densities disagreed with one another by at least a factor of two. Based on our results, we believe this difference is due to theoretical uncertainties in the $(186.85 + 186.88)/195.12$ diagnostic. 
 
\subsection{Fe~\textsc{xiii}}\label{subsec:FeXIII}

We were able to identify one density-insensitive Fe~\textsc{xiii} line ratio involving the $3s^2\,3p^2\:^{3}P_{0}-3s^2\,3p\,3d\:\:^{3}D^{o}_{1}$ and $3s^2\,3p^2\:^{3}P_{2} - 3s^2\,3p\,3d\:\:^{3}D^{o}_{1}$ transitions at 197.43 and 204.93~\AA, respectively \citep{DelZanna:AA:2012b}. These lines are relatively weak compared to the other spectral lines that we are interested in, but they could be measured reliably for higher electron beam currents. Figure~\ref{fig:ratio3} plots the measured ratios for these lines and compares them to the predictions of FAC. The ratio is indeed density insensitive. However, we found that the ratio disagreed with the FAC calculations. The weighted mean of our measurements is $0.53 \pm 0.01$. 

Initially, the theoretical calculations were performed with limited
configuration interaction that included only states with one electrons in the
$3d$ orbital, and produced a predicted ratio of $\sim$ 1. We recalculated the
ratio by including configurations with double excitation to the $3d$ orbitals.
This reduced the predicted ratio to $\sim$ 0.82, and is close to the value in
CHIANTI of about 0.86. Clearly, a large discrepancy exists between theory and
experiment. These two lines originate from the same upper level, their
intensity ratio is therefore strictly that of the respective radiative decay
rates, and depend solely on the atomic structure model. It is possible that
including even more configuration interaction in the model might resolve the
discrepancy. However, including all double excitations to even higher orbitals
results in extremely large Hamiltonian matrices, and the model quickly
becomes computationally intractable. A more manageable solution is to use a
many-body perturbation approach to evaluate the effect of further electron
correlation effects. However, this is beyond the scope of the present study.


The intensity ratio of the Fe~\textsc{xiii} lines within the 203.83~\AA\ line blend (see Table~\ref{table:table1}) and the 202.04~\AA\ line is plotted in Figure \ref{fig:ratio4}. For the densities accessible in the experiment, we are in the high density limit of this diagnostic and observed little variation in the ratio as a function of density. The ratio itself was higher by about 25\% than that predicted by the FAC calculations, though as we explain below this appears to lie within the combined experimental and theoretical uncertainties. The same line ratio in CHIANTI for a $T=4 \times 10^{6}$~K Maxwellian agrees with our FAC calculations to within a few percent. Hence, the difference we find with experiment also exists in the CHIANTI data. 

We considered the effects of polarization as a possible explanation for this difference, but we find that polarization cannot account for the difference between theory and experiment. According to \citet{Liang:ApJ:2009a} the polarization of most of the Fe~\textsc{xiii} lines that we have measured is only a few percent and the effects on intensity are negligible. The Fe~\textsc{xiii} 202.04~\AA\ line, though, has a significant polarization of 21\%. Spiraling of the electron beam reduces this to less than 18\%, which is predicted to cause at most a 6\% difference in the measured versus theoretical intensity. However, this polarization effect would imply that we have overestimated the intensity of the $202.04$~\AA\ line. As this line is in the denominator of our density diagnostic, correcting for this effect would actually exacerbate the difference with theory, rather than resolve it. 

Blending issues in the experiment are another possible explanation for the disagreement with theory. The 203.83~\AA\ lines form a very complex blend, and it is possible that there is a systematic overestimate of its intensity. The complex consists of six closely spaced lines, four that are identified as Fe~\textsc{XII} 203.72 \AA\ and Fe~\textsc{XIII} 203.77, 203.80, and 203.83~\AA, plus two other unidentified lines at 203.60 and 203.66~\AA. Further, we have observed O~\textsc{V} lines in this spectrum, therefore it is possible that the weak O V $1s^2\,2p^2\:{^3}{P}{_{0}}$ -  $1s^2\,2p(2P^o_{1/2})3d\:{^3}{D}{^o_{1}}$ and $1s^2\,2p^2\:{^3}{P}{_{1}}$ -  $1s^2\,2p(2P^o_{1/2})3d\:{^3}{D}{^o_{1}}$ transitions at 203.78 and 203.82~\AA\ may also contribute to this complex. Given these experimental factors and the estimated 20\% systematic uncertainty in the calculations, we estimate that the observed differences between theory and experiment for this ratio are within the combined uncertainties. 

In Figure~\ref{fig:ratio4} we also plot the intensity ratios for the 196.63/202.04 and 200.02/202.04 density diagnostics. For the 196.53/202.04 ratio, the density diagnostic is not quite saturated and we find that the experimental ratio is about 20\% larger than the theoretical prediction, but within the combined experimental and theoretical uncertainties. Our measurements of the 200.02/202.04 diagnostic are in the high-density saturation limit of the diagnostic and the experimental ratios are in good agremeent with the calculations. 

 
\subsection{Fe~\textsc{xiv}}\label{subsec:Fe-XIV}
 
We have also measured density-sensitive Fe~\textsc{xiv} lines in the wavelength range 255--277~\AA\ \citep{Malinovsky:ApJ:1973, DelZanna:AA:2012}. The spectra in this range are much cleaner, with fewer blends than in the short wavelength range containing the Fe~\textsc{xii} and \textsc{xiii} diagnostics. However, the intensities of the lines were somewhat weaker and required longer exposure times. Because our results for Fe~\textsc{xii} and \textsc{xiii} found that there was no line-ratio dependence seen between $E_{\mathrm{e}}=395$ versus 475~eV, we carried out experiments only for $E_{\mathrm{e}}=395$~eV, enabling us to obtain longer exposure spectra. 

We measured the line-intensity ratios for 257.39/274.21, 264.79/274.21, and 270.52/274.21. The specific transitions are listed in Table~\ref{table:table2}. Our results are shown in Figure~\ref{fig:ratio5}, where they are compared to the FAC calculations. For these ratios, we found reasonable agreement between these measured and predicted ratios. 
According to \citet{DelZanna:AA:2012}, the Fe~\textsc{xiv} line at 264.79~\AA\ is blended with an Fe~\textsc{xi} spectral line, though the precise wavelength of that line is unknown. A blend contribution to the 264.79~\AA\ line would systematically increase the measured line ratio relative to the prediction, which currently appears to be in excellent agreement with the experimental results. It is possible that the blend is present in solar spectra, but negligible in the experiment. However, we observe two Fe~\textsc{x} lines in the spectrum (see Figure~\ref{fig:spec1}), which suggest that the Fe~\textsc{xi} line should also be present in our spectrum. The fact that we do not notice any blending seems to indicate that the Fe~\textsc{xi} line is either not at the predicted location or it is weaker than calculated. 





\subsection{Discussion of the Results}

Our measurements have identified several discrepancies with theoretical calculations. For Fe~\textsc{xii}, the (186.85 + 186.88)/195.12 line ratio does not match the theoretical calculations. In particular, the measured ratio is larger, has a steeper density dependence and has a greater value for the ratio in the high density limit. A similar discrepancy exists for the ratio of the 186~\AA\ lines with either the 193.51 or 192.39~\AA\ lines, as we have confirmed that the 192.39, 193.51, and 195.12~\AA\ lines are density insensitive with respect to one another.

For Fe~\textsc{xiii} a significant discrepancy was found for the density-insensitive ratio of the 197.43/204.94 lines. Our measurements confirm that the ratio is not sensitive to density, but the magnitude of the ratio differs by $\approx30\%$ from the value predicted by FAC or CHIANTI.

We measured three Fe~\textsc{xiii} density diagnostics. Of these, the 200.02/202.04 line ratio appears to be the most accurate. Reasonable agreement between experiment and theory is also found for the 196.52/202.04 line ratio. For the (203.77 + 203.79 + 203.83)/202.04 diagnostic, our measurements are all in the high density saturation limit of the diagnostic, where we found the measured ratio to be somewhat larger than predicted by theory. This diagnostic involves a complex blend that adds to the uncertainty in the measurement so that the disagreement is at the limit of the combined theoretical and experimental uncertainties. 

There are two other EBIT measurements of this ratio. \citet{Liang:ApJ:2009a} measured this ratio at a density of $\approx 3\times10^{9}$~cm$^{-3}$ and their measured value is in good agreement with the calculation. \citet{Yamamoto:ApJ:2008} reported a measurement of this ratio from the LLNL EBIT-II electron beam ion trap and again obtained good agreement at $\approx 2 \times 10^{11}$~cm$^{-3}$. Two measurements using magnetic fusion devices with densities near $2\times10^{13}$~cm$^{-3}$ were also made using the Large Helical Device \citep{Yamamoto:ApJ:2008} and the National Spherical Torus Experiment \citep{Weller:ApJ:2018}. Both of these produced values well below theory. However, they used spectrometers that did not resolve the extensive line blending with emission from neighboring charge states of iron. 


Our measurements for several Fe~\textsc{xiv} line ratios were all in agreement with theoretical calculations. Though, again, our densities were near or above the high density saturation of those diagnostics. The 264.79/274.21 and 270.52/274.21 diagnostics were also measured at lower densities by \citet{Nakamura:ApJ:2011}, who found good agreement with atomic calculations. Based on our results and those of \citeauthor{Nakamura:ApJ:2011}, these Fe~\textsc{xiv} diagnostics appear to be reliable.

Measurements on the National Spherical Torus Experiment at $2\times10^{13}$~cm$^{-3}$ \citep{Weller:ApJ:2018} agree well with the prediction for the 270.52/274.21 ratio, but are significantly larger than predicted for the 264.79/274.21 ratio. Although we do not expect line blending, these measurements were performed with much lower spectral resolution and blending cannot be ruled out. Blending is also more likely in the tokamak experiments as those plasmas contained multiple other elements besides iron. 

\section{Summary} \label{sec:summary}

We have measured density-diagnostic line-intensity ratios for EUV lines from Fe~\textsc{xii}, \textsc{xiii}, and \textsc{xiv} using an EBIT. In our analysis, we identified several discrepancies between our measurements and the existing theory calculations. Of those, the most glaring are the commonly used Fe~\textsc{xii} (186.85 + 186.88)/195.12 line ratio and the density-insensitive Fe~\textsc{xiii} 197.43/204.94 line ratio. The latter is not a density diagnostic, but it is of concern because it might suggest other possible problems with the Fe~\textsc{xiii} atomic model. On the other hand, the Fe~\textsc{xiii} density diagnostics that we studied did not exhibit such large discrepancies. 

A positive result is that we find that density diagnostics using the Fe~\textsc{xiii} 196.53/202.04 and Fe~\textsc{xiv} 264.79/274.21, and 270.52A/274.21 line ratios are reliable using the FAC atomic data. With regards to the discrepancies noted by \citet{Young:AA:2009} between the Fe~\textsc{xii} (186.85 + 186.88)/195.12 and Fe~\textsc{xiii} 196.54/202.04 diagnostics, our results suggest that the Fe~\textsc{xiii} diagnostic is likely reliable and that the problem lies with the Fe~\textsc{xii} diagnostic, with the caveat that we are comparing to FAC monoenergetic beam calculations, but these data appear not to differ significantly from the data tabulated in CHIANTI that were used by \citet{Young:AA:2009}.
 
\acknowledgements

This work is supported, in part, by NASA H-TIDeS grant NNX16AF10G. Work at LLNL is performed under the auspices of the U.S. DOE under contract No. DE-AC52-07NA27344.

\appendix

\section{Derivation of the Effective Density}\label{app:appendix1}
Here we present a brief derivation of the spatially averaged density of Equation~(\ref{eq:neff}) in order to clarify what is represented by that quantity and then extend that result to derive Equation~(\ref{eq:neffdbl}) for a double-Gaussian ion cloud. In each case, the electron density distribution is taken to be a Gaussian given by Equation~(\ref{eq:negauss}). 
For a Gaussian ion cloud, the fraction of ions in a radius between $r$ and $r + \diff{r}$ is given by 
\begin{equation}
f_{\mathrm{i}}(r) = \frac{1}{2 \pi \sigma_{\mathrm{i}}^{2}} e^{-r^2/2\sigma_{\mathrm{i}}^{2}}, 
\label{eq:fi}
\end{equation}
where $\sigma_{\mathrm{i}}$ is the 1$\sigma$ radius of the Gaussian ion cloud. The normalization constant has been chosen so that the integral of $f_{\mathrm{i}}(r)$ is unity. 

The spatially averaged electron density, $n_{\mathrm{eff}}$, is the integral of the fraction of the ions at a given radius multiplied by the electron density at that radius. That is, 
\begin{equation}
n_{\mathrm{eff}} = \int_{0}^{2\pi} \int_{0}^{\infty} f_{\mathrm{i}}(r) n_{\mathrm{e}}(r) r \diff{r} \diff{\theta}
\label{eq:neff1a}
\end{equation}
Substituting in Equation~(\ref{eq:negauss}) and Equation~(\ref{eq:fi}), we find
\begin{equation}
n_{\mathrm{eff}} = n_0 \frac{\sigma_{\mathrm{e}}^2}{\sigma_{\mathrm{e}}^2+\sigma_{\mathrm{i}}^2}
\label{eq:neff1}
\end{equation}
In order to express this in the same terms as Equation~(\ref{eq:neff}), we use the fact that the electron current is related to the central density by
\begin{equation}
I_{\mathrm{e}} = 2 \pi e v_{\mathrm{e}} n_{0} \sigma_{\mathrm{e}}^2,   
\label{eq:necentral}
\end{equation}
and that the FWHM of a Gaussian is $\Gamma = 2\sigma\sqrt{2\ln{2}}$. Substituting these expressions into Equation~(\ref{eq:neff1}) produces Equation~(\ref{eq:neff}).

The double Gaussian ion cloud that we measure on the detector is parameterized as
\begin{equation}
F_{\mathrm{i}}(x) = A_1 e^{-x^2/2 \sigma_1^2} + A_2 e^{-x^2/2 \sigma_2^2},
\label{eq:dblproject}
\end{equation}
where the coordinate $x$ is the projection of the cloud radius onto the detector. The radial distribution of the ions is then given by the Abel transform pair
\begin{equation}
f_{\mathrm{i}}(r) = \frac{A_1}{\sigma_1 \sqrt{2\pi}} e^{-r^2/2 \sigma_1^2} + \frac{A_2}{\sigma_2 \sqrt{2\pi}} e^{-r^2/2 \sigma_2^2}.
\label{eq:dblr}
\end{equation}
One finds Equation~(\ref{eq:neffdbl}) upon carrying out the integral of Equation~(\ref{eq:neff1a}), then changing variables using Equation~(\ref{eq:necentral}), and using the relation between $\sigma$ and $\Gamma$.

\section{Equivalence of the Spatially Averaged  Density and Time-Averaged Density}\label{app:appendix2}

As mentioned in Section~\ref{sec:effdensity}, it is the time-averaged density experienced by the ions, rather than the spatially averaged density, $n_{\mathrm{eff}}$, that is relevant for determining the ion populations. This is because the process of excitation and relaxation occur as functions of time. Here, we show that the spatially averaged effective density is equivalent to the time-averaged density experienced by the collection of ions in the trap, as long as the trajectories of the ions are uncorrelated. 

For an ion in EBIT the density is a function only of the radial position of the ion. Consider breaking the ion trajectories up into discrete time steps. For each ion, we denote the radial position at the $j$-th time step as $r(t_j)$. Also, we label each of a large number of ions by the index $i$, so that the electron density experienced by the $i$-th ion at time $t_j$ along its trajectory is $n_{i}(r(t_j))$. For clarity, we have dropped the label ``e'' indicating electron density, as all the densities described here are electron densities.

The time-averaged electron density experienced by an individual ion is, 
\begin{equation}
\avg{n_{i}} = \frac{1}{N_{T}} \sum_j n_{i}(r(t_j)), 
\label{eq:tavg}
\end{equation}
where $N_{T}$ is the total number of time steps. However, we are not interested in the time-average for any particular ion, but rather in the time-averaged density experienced by all the ions in the trap, or at least a representative sample of those ions. This average over the ions of the time-averaged density experienced by each individual ion is given by: 
\begin{equation}
\avg{n} = \frac{1}{N_{I}} \sum_i \avg{n_{i}},
\label{eq:xavg}
\end{equation}
where the average is taken over a sample of $N_{I}$ ions. 

Now, we substitute Equation~(\ref{eq:tavg}) into Equation~(\ref{eq:xavg}) and switch the order of the summations. 
Then we have, 
\begin{equation}
\avg{n} = \frac{1}{N_{T}} \sum_{j} \frac{1}{N_{I}} \sum_{i} n_{i}(r(t_{j}))
\label{eq:txavg}
\end{equation}
The inner summation over $i$ is the average density experienced by the ions at a fixed time. As long as the ions are randomly formed in space, they are not biased in how they sample the electron density throughout the ion cloud. Hence, the average electron density sampled by the ions is equivalent to the spatial average of the electron density, $n_{\mathrm{eff}}$. 

Moreover, if the ions are randomly distributed at any time-step and their trajectories are uncorrelated, then this spatial average does not vary in time. Consequently, we do not need to take the time average at all, because the time-average is just the average of a constant value. So, we can use
\begin{equation}
\avg{n} = n_{\mathrm{eff}} = \frac{1}{N_{I}} \sum_{i} n_{i}(r(t_j)). 
\label{eq:simpleavg}
\end{equation}
This proves that the spatial average is equivalent to the time average. 

\newpage
\clearpage

\newpage
\clearpage

\newpage
\clearpage

\newpage
\clearpage

\newpage
\clearpage

\newpage
\clearpage

\newpage
\clearpage

\newpage
\clearpage

\newpage
\clearpage

\newpage
\clearpage

%

\newpage
\clearpage

\startlongtable
\begin{deluxetable}{ccccc}
	\tablewidth{0pt}
	\tablecolumns{6}
	\tablecaption{\label{table:table1} Observed spectral lines of interest in Figure~\ref{fig:spec1}.}
	\tablehead{
		\colhead{Wavelength (\AA)} & 
		\colhead{Ion\tablenotemark{a}} & 
		\colhead{Lower Level} & 
		\colhead{Upper Level} & 
		\colhead{Comments}
	}
\startdata
$185.21$ & Fe  VIII & $3p^63d \>\> {^2}{D}{_{5/2}}$ & $3p^5(^2P^o)3d^2(^3F) \>\> {^2}{F}{^o_{7/2}}$  & \\
$186.60$ & Fe VIII & $3p^63d \>\> {^2}{D}{_{3/2}}$ & $3p^5(^2P^o)3d^2(^3F) \>\> {^2}{F}{^o_{5/2}}$  & \\
$186.75$ & U & \nodata & \nodata& Blended\\
$186.80$ & U & \nodata & \nodata & \ditto \\
$186.85$ & Fe  XII & $3s^23p^3 \>\> {^2}{D}{^o_{3/2}}$ & $3s^23p^2(^2P)3d \>\> {^2}{F}{_{5/2}}$  & \ditto \\
$186.88$ & Fe XII & $3s^23p^3 \>\> {^2}{D}{^o_{5/2}}$ & $3s^23p^2(^2P)3d \>\> {^2}{F}{_{7/2}}$  & \ditto \\
$188.19$ & Fe  XII & $3s^23p^3 \>\> {^2}{P}{^o_{1/2}}$ & $3s^23p^2(^2P)3d \>\> {^2}{D}{_{3/2}}$  & Blended \\
$188.21$ & Fe  XI & $3s^23p^4 \>\> {^3}{P}{_{2}}$ & $3s^23p^3(^2D^2)3d \>\> {^3}{P}{^o_{2}}$  & \ditto \\
$188.30$ & Fe  XI & $3s^23p^4 \>\> {^3}{P}{_{2}}$ & $3s^23p^3(^2D)3d \>\> {^3}{P}{_{2}}$  & \\
$190.98$ & U & \nodata& \nodata& Blended\\
$191.04$ & Fe  XII & $3s^23p^3 \>\> {^2}{P}{^o_{3/2}}$ & $3s^23p^2(^3P)3d \>\> {^2}{D}{_{5/2}}$  &\ditto \\
$192.02$ & Fe  XI & $3s^23p^4 \>\> {^3}{P}{_{1}}$ & $3s^23p^3(^2D^o)3d \>\> {^1}{P}{^o_{1}}$  &\\
$192.29$ & U & \nodata& \nodata& Blended\\
$192.34$ & U & \nodata& \nodata& \ditto \\
$192.39$ & Fe  XII & $3s^23p^3 \>\> {^4}{S}{^o_{3/2}}$ & $3s^23p^2(^3P)3d \>\> {^4}{P}{_{1/2}}$  & \ditto \\
$192.63$ & Fe  XIV & $3s3p^2 \>\> {^2}{D}{_{5/2}}$ & $3s3p^3(^1P^o)3d \>\> {^2}{F}{^o_{5/2}}$  &\\
$192.75$ & O V & $1s^22s2p \>\> {^3}{P}{^o_{0}}$ &  $1s^22s3d \>\> {^3}{D}{_{1}}$  & Blended\\
$192.82$ & Fe  XI & $3s^23p^4 \>\> {^3}{P}{_{1}}$ & $3s^23p^3(^2D^0)3d \>\> {^3}{P}{^o_{2}}$  & \ditto \\
$192.91$ & O V & $1s^22s2p \>\> {^3}{P}{^o_{2}}$ &  $1s^22s3d \>\> {^3}{D}{_{3}}$  &\ditto \\
$193.42$ & Fe VII & $3p^63d^2 \>\> {^3}{P}{_{2}}$ & $3p^5(^2P^o)3d^3(^2D^o_2) \>\> {^1}{D}{_{2}}$  & Blended\\
$193.46$ & U & \nodata& \nodata& \ditto \\
$193.51$ & Fe XII & $3s^23p^3 \>\> {^4}{S}{^o_{3/2}}$ & $3s^23p^2(^3P)3d \>\> {^4}{P}{_{3/2}}$  & \ditto \\
$195.02$ & U & \nodata& \nodata& Blended\\
$195.06$ & U & \nodata& \nodata& \ditto \\
$195.12$ & Fe XII & $3s^23p^3 \>\> {^4}{S}{^o_{3/2}}$ & $3s^23p^2(^3P)3d \>\> {^4}{P}{_{5/2}}$  & \ditto \\
$195.18$ & Fe XII & $3s^23p^3 \>\> {^2}{D}{_{3/2}}$ & $3s^23p^2(^1D)3d \>\> {^2}{D}{_{3/2}}$  & \ditto \\
$196.42$ & Fe VII & $3p^63d^2 \>\> {^3}{F}{_{2}}$ & $3p^5(^2P^o)3d^3(^2H) \>\> {^3}{G}{^o_{4}}$  & Blended\\
$195.47$ & U & \nodata& \nodata& \ditto \\
$196.53$ & Fe  XIII & $3s^23p^2 \>\> {^1}{D}{_{2}}$ & $3s^23p3d \>\> {^1}{F}{^o_{5/2}}$  & \ditto \\
$196.64$ & Fe  XII & $3s^23p^3 \>\> {^2}{D}{^o_{5/2}}$ & $3s^23p^2(^1D)3d \>\> {^2}{D}{_{5/2}}$  & \ditto  \\
$196.92$ & Fe XII & $3s^23p^3 \>\> {^2}{D}{^o_{5/2}}$ & $3s^23p^2(^1D)3d \>\> {^2}{D}{_{3/2}}$  & \\
$197.02$ & C V & $1s2p \>\> {^1}{P}{^o_{1}}$ & $1s4d \>\> {^1}{D}{_{2}}$  & \\
$197.43$ & Fe  XIII & $3s^23p^2 \>\> {^3}{P}{_{0}}$ & $3s^23p3d \>\> {^3}{D}{^o_{1}}$  &\\
$197.86$ & Fe IX & $3s^23p^53d \>\> {^1}{P}{_{1}}$ & $3s^23p^54p \>\> {^1}{S}{_{0}}$  & \\
$199.90$ & U & \nodata& \nodata& Blended\\
$199.96$ & U & \nodata& \nodata& \ditto \\
$200.02$ & Fe  XIII & $3s^23p^2 \>\> {^3}{P}{_{1}}$ & $3s^23p3d \>\> {^3}{D}{^o_{2}}$  & \ditto \\
$200.36$ & Fe XII & $3s^23p^3 \>\> {^2}{P}{^o_{3/2}}$ & $3s^23p^2(^1D)3d \>\> {^2}{S}{_{1/2}}$  & \\
$201.11$ & Fe  XI & $3s^23p^4 \>\> {^3}{P}{_{2}}$ & $3s^23p^3(^2P)3d \>\> {^3}{D}{_{3}}$  & Blended\\
$201.13$ & Fe XIII & $3s^23p^2 \>\> {^3}{P}{_{1}}$ & $3s^23p3d \>\> {^3}{D}{^o_{1}}$  & \ditto \\
$201.14$ & Fe XII & $3s^23p^3 \>\> {^2}{P}{^o_{3/2}}$ & $3s^23p^2(^3P)3d \>\> {^2}{P}{_{3/2}}$  &  \ditto \\
$201.73$ & Fe  XI & $3s^23p^4 \>\> {^1}{D}{_{2}}$ & $3s^23p^3(^2D^o)3d \>\> {^1}{P}{^o_{1}}$  & Blended \\
$201.74$ & Fe XII & $3s^23p^3 \>\> {^2}{P}{_{1/2}}$ & $3s^23p^23d\>\> {^2}{P}{_{1/2}}$ &  \ditto \\
$201.76$ & Fe XII & $3s^23p^3 \>\> {^2}{P}{_{3/2}}$ & $3s^23p^23d\>\> {^2}{S}{_{1/2}}$ &  \ditto \\
$201.77$ & U & \nodata& \nodata& \ditto \\
$201.92$ & U & \nodata& \nodata& Blended\\
$201.98$ & U & \nodata& \nodata& \ditto \\
$202.04$ & Fe XIII & $3s^23p^2 \>\> {^3}{P}{_{0}}$ & $3s^23p3d \>\> {^3}{P}{^o_{1}}$  & \ditto \\
$203.17$ & Fe XIII & $3s^23p^2 \>\> {^3}{P}{_{1}}$ & $3s^23p3d \>\> {^3}{P}{_{0}}$  & \\
$203.61$ & U & \nodata& \nodata& Blended\\
$203.66$ & U & \nodata& \nodata& \ditto \\
$203.73$ & Fe XII & $3s^23p^3 \>\> {^2}{D}{_{5/2}}$ & $3s^23p^2(^1S)3d \>\> {^2}{D}{_{5/2}}$  & \ditto  \\
$203.77$ & Fe XIII & $3s^23p^2 \>\> {^3}{D}{_{1}}$ & $3s^23p3d \>\> {^3}{F}{_{2}}$  & \ditto \\
$203.79$ & Fe  XIII & $3s^23p^2 \>\> {^3}{P}{_{2}}$ & $3s^23p3d \>\> {^3}{D}{^o_{2}}$  & \ditto \\
$203.83$ & Fe  XIII & $3s^23p^2 \>\> {^3}{P}{_{2}}$ & $3s^23p3d \>\> {^3}{D}{^o_{3}}$  & \ditto \\
$204.26$ & Fe XIII & $3s^23p^2 \>\> {^3}{P}{_{1}}$ & $3s^23p3d \>\> {^1}{D}{^o_{2}}$  & \\
$204.93$ & Fe XIII & $3s^23p^2 \>\> {^3}{P}{_{2}}$ & $3s^23p3d \>\> {^3}{D}{^o_{1}}$  & \\
\enddata
\tablenotetext{a}{U-Unidentified spectral line}	
\tablecomments{Wavelengths and transitions are from CHIANTI \citep{Dere:AAS:1997, Landi:ApJ:2013, DelZanna:AA:2015}.}
\end{deluxetable}

 \newpage
\clearpage

\begin{deluxetable}{ccccc}
	\tablewidth{0pt}
	\tablecolumns{6}
	\tablecaption{\label{table:table2} Observed spectral lines of interest in Figure~\ref{fig:spec2}.}
	\tablehead{
		\colhead{Wavelength (\AA)} & 
		\colhead{Ion} & 
		\colhead{Lower Level} & 
		\colhead{Upper Level} & 
		\colhead{comments}
	}
	\startdata
$256.38$ & Fe X & $3s^23p^5 \>\> {^2}{P}{^o_{3/2}}$ & $3s^23p^4(^3P)3d \>\> {^4}{D}{_{3/2}}$  & Blended\\
$256.42$ & Fe XIII & $3s^23p^2 \>\> {^1}{D}{_{2}}$ & $3s3p^3 \>\> {^1}{P}{^o_{1}}$  & \ditto \\
$257.26$ & Fe  X & $3s^23p^5 \>\> {^2}{P}{^o_{3/2}}$ & $3s^23p^4(^3P)3d \>\> {^4}{D}{_{7/2}}$  & Blended \\
$257.39$ & Fe  XIV & $3s^23p \>\> {^2}{P}{^o_{1/2}}$ & $3s3p^2 \>\> {^2}{P}{_{1/2}}$  &  \ditto \\
$264.79$ & Fe  XIV & $3s^23p \>\> {^2}{P}{^o_{3/2}}$ & $3s3p^2 \>\> {^2}{P}{_{3/2}}$  & \\
$270.51$ & Fe  XIV & $3s^23p \>\> {^2}{P}{^o_{3/2}}$ & $3s3p^2 \>\> {^2}{P}{_{1/2}}$  & \\
$274.20$ & Fe  XIV & $3s^23p \>\> {^2}{P}{^o_{1/2}}$ & $3s3p^2 \>\> {^2}{S}{_{1/2}}$  & \\\hline
\enddata
\tablecomments{Wavelengths and transitions are from CHIANTI \citep{Dere:AAS:1997, Landi:ApJ:2013, DelZanna:AA:2015}.}
\end{deluxetable}


\newpage
\clearpage

\begin{table}[h]
	\begin{center}
		\caption{\label{table:table3} Measured line intensities for $E_{\mathrm{e}} = 395$~eV in the 185--205~\AA\ range versus electron beam current.}
		\begin{tabular}{ c c c c c c c }
			\hline
			Spectral Line(s) & \multicolumn{6}{c}{Spectral Line Intensity }\\ 
			
			(\AA) & \multicolumn{6}{c} {(photon number)} \\
			\cline{2-7}
			&1~mA  & 2~mA  & 3~mA  & 5~mA  & 7~mA  & 8~mA   \\
			\hline
			186.85+186.88 & 11561 & 15524 & 20188 & 21913 & 29977 & 10203 \\
			& $\pm$ $\pm$ 86 & $\pm$ 119 & $\pm$ 250 & $\pm$ 263 & $\pm$ 171 & $\pm$ 103 \\ 
			192.39 & 3094 & 3126 & 3706 & 3888 & 5168 & 1689 \\
			& $\pm$ 56 &  $\pm$ 54 &  $\pm$ 62 &  $\pm$ 63 &  $\pm$ 73 &  $\pm$ 42 \\ 
			193.51 &  5815 & 6898 & 8146 & 8420 & 11228 & 3518 \\
			& $\pm$ 78 &  $\pm$ 81 &  $\pm$ 91 &  $\pm$ 93 &  $\pm$ 108 &  $\pm$ 61 \\ 
			195.12 &  9623 & 10068 & 11968 & 12311 & 16728 & 5346  \\
			& $\pm$ 96 &  $\pm$ 99 &  $\pm$ 111 &  $\pm$ 111 &  $\pm$ 129 &  $\pm$ 76 \\ 
			196.53 & 3733 & 5208 & 7883 & 91 & 14584 & 5201 \\
			& $\pm$ 63 &  $\pm$ 71 &  $\pm$ 91 &  $\pm$ 98 &  $\pm$ 123 &  $\pm$ 76 \\ 
			196.64 &  2089 & 2608 & 3204 & 3829 & 5058 & 1532 \\
			& $\pm$ 47 &  $\pm$ 51 &  $\pm$ 58 &  $\pm$ 63 &  $\pm$ 72 &  $\pm$ 42 \\ 
			197.43 &   &  & 500.63 & 708 & 964 & 366 \\
			&  &   &  $\pm$ 22 &  $\pm$ 29 &  $\pm $ 36 &  $\pm $ 22 \\
			200.02&  1422 & 2118 & 3064 & 3464 & 5291 & 1984 \\
			& $\pm$ 34 &  $\pm$ 43 &  $\pm$ 60 &  $\pm$ 60 &  $\pm$ 73 &  $\pm$ 47 \\
			202.04& 2061 & 3162 & 3893 & 4663 & 6748 & 2246 \\
			& $\pm$ 42 &  $\pm$ 59 &  $\pm$ 64 &  $\pm$ 71 &  $\pm$ 73 &  $\pm$ 52 \\
			203.77+203.79+203.83& 10464 & 16257 & 19969 & 23619 & 33181 & 11613 \\
			& $\pm$ 109 &  $\pm$ 139 &  $\pm$ 162 &  $\pm$ 168 &  $\pm$ 196 &  $\pm$ 124 \\
			204.94&  &  & 967.5 & 1108 & 1799 & 731 \\
			&   &    &  $\pm$ 36 &  $\pm$ 38 &  $\pm$ 49 &  $\pm$ 31 \\
			\hline
		\end{tabular}
	\end{center}
\end{table}

\newpage
\clearpage

\begin{table}[h]
	\begin{center}
		\caption{\label{table:table4} Same as Table~\ref{table:table3}, but for $E_{\mathrm{e}}=475$~eV.}
		\begin{tabular}{c c c c c c c }
			\hline
			Spectral Line(s) & \multicolumn{6}{c}{Spectral Line Intensity }\\ 
			
			(\AA) & \multicolumn{6}{c} {(photon number)}\\
			\cline{2-7}
			&1~mA  & 2~mA  & 3~mA  & 5~mA  & 7~mA  & 9~mA   \\
			\hline
			186.85+186.88 & 5899 & 3093 & 16753 & 21328 & 18235 & 21396 \\
			& $\pm$ 73  &  $\pm$ 53  &  $\pm$ 231 &  $\pm$ 258 &  $\pm$ 133  &  $\pm$ 149 \\ 
			192.39 & 1168 & 696 & 3209 & 3793 & 3389 & 4712 \\
			& $\pm$ 31  &  $\pm$ 26  &  $\pm$ 57  &  $\pm$ 64 &  $\pm$ 59 &  $\pm$ 70 \\ 
			193.51 &  2596 & 1606 & 7178 & 8664 & 6768 & 9405 \\
			& $\pm$ 46 &  $\pm$ 39  &  $\pm$ 86  &  $\pm$ 97  &  $\pm$ 83  &  $\pm$ 99 \\ 
			195.12 &  4452 & 2036 & 10250 & 11430 & 9518 & 11230 \\
			& $\pm$ 63 &  $\pm$ 46  &  $\pm$ 100  &  $\pm$ 106  &  $\pm$ 94  &  $\pm$ 106  \\ 
			196.53 & 2553 & 1294 & 6299 & 8812 & 7649 & 10783 \\
			& $\pm$ 49 &  $\pm$ 36  &  $\pm$ 78  &  $\pm$ 96  &  $\pm$ 88 &  $\pm$ 106 \\ 
			196.64 &  1161 & 649 & 2951 & 3424 & 3248 & 4672 \\
			& $\pm$ 33  &  $\pm$ 25  &  $\pm$ 54  &  $\pm$  60 &  $\pm$ 57 &  $\pm$ 69  \\ 
			197.43 &   &  &  & 458 & 389 & 617 \\
			&  &   &   &  $\pm$ 22 &  $\pm $ 21 &  $\pm $ 26  \\
			200.02&  1201 & 546 & 3271 & 3688 & 2763 & 4257 \\
			& $\pm$ 31 &  $\pm$ 25  &  $\pm$ 68  &  $\pm$ 63  &  $\pm$ 52 &  $\pm$ 65  \\
			202.04& 1445 & 763 & 3564 & 4488 & 3568 & 5443 \\
			& $\pm$ 36 &  $\pm$ 30 &  $\pm$ 59  &  $\pm$ 69 &  $\pm$ 59 &  $\pm$ 75 \\
			203.77+203.79+203.83& 7266 & 3531 & 17193 & 21373 & 17675 & 24656 \\
			& $\pm$ 90 &  $\pm$ 63  &  $\pm$ 138  &  $\pm$ 145 &  $\pm$ 137 &  $\pm$ 173 \\
			204.94&  &  &  & 917 & 854 & 1086\\
			&   &    &   &  $\pm$ 32 &  $\pm$ 33 &  $\pm$ 36  \\
			\hline
			
		\end{tabular}
	\end{center}
\end{table}


\newpage
\clearpage


\begin{table}[h]
	\begin{center}
		\caption{\label{table:table5} Same as Table~\ref{table:table3}, but in the 255--276~\AA\ range.} 
		\begin{tabular}{c c c c c}
			\hline
			Spectral Line & \multicolumn{4}{c}{Spectral Line Intensity}\\ 
			(\AA) & \multicolumn{4}{c}{(photon number)}\\
			\cline{2-5}
			&3~mA  & 5~mA  & 7~mA  & 8~mA   \\
			\hline
			257.39 & 2288 & 3272 & 4487 & 3249 \\
			& $\pm$ 39  &  $\pm$ 55  &  $\pm$ 65 &  $\pm$ 47  \\ 
			264.79 &  8913 & 14026 & 16147 & 10797\\
			& $\pm$ 91  &  $\pm$ 115  &  $\pm$ 126  &  $\pm$ 86  \\ 
			270.52 &  2972 & 4948 & 5585 & 4221 \\
			& $\pm$ 50 &  $\pm$ 69  &  $\pm$ 72  &  $\pm$ 54   \\ 
			274.21 &  3708 & 5382 & 6305 & 4638 \\
			& $\pm$ 55 &  $\pm$ 70  &  $\pm$ 78  &  $\pm$ 56  \\ 
			\hline
		\end{tabular}
	\end{center}
\end{table}

\newpage
\clearpage

\begin{deluxetable}{CCCCCCCC}
	\tablewidth{0pt}
	\tablecolumns{8}
	\tablecaption{\label{table:table6} Electron cloud width, $\Gamma_{\mathrm{e}}$; ion cloud component amplitudes $A_{1,2}$, and widths $\Gamma_{1,2}$; Nominal beam density, $\bar{n}_{\mathrm{e}}$; and effective density, $n_{\mathrm{eff}}$, for $E_{\mathrm{e}}=395$~eV.}
	\tablehead{
		\colhead{Beam current} & 
		\colhead{$\Gamma_{\mathrm{e}}$} & 
		\colhead{$A_1$} &
		\colhead{$\Gamma_1$} & 
		\colhead{$A_2$} & 
		\colhead{$\Gamma_2$} & 
		\colhead{$\bar{n}_{\mathrm{e}}$} & 
		\colhead{$n_{\mathrm{eff}}$} \\
		\colhead{(mA)} & 
		\colhead{$\mathrm{(\mu m)}$} & 
		\colhead{(counts)} & 
		\colhead{$\mathrm{(\mu m)}$} & 
		\colhead{(counts)} & 
		\colhead{$\mathrm{(\mu m)}$} & 
		\colhead{$(10^{11}\: \mathrm{cm^{-3}})$} & 
		\colhead{$(10^{11}\: \mathrm{cm^{-3}})$}
	}
	\startdata
	1 & 56.4 \pm 7.4 & 1958 \pm 88 & 114.6 \pm 5.6 & 0 & 0 & 0.74 & 0.288 \pm 0.027 \\
	2 & 55.6 \pm 7.1 & 4547 \pm 157  & 93.5 \pm 3.8 & 545 \pm 127 & 392 \pm 84 & 1.53 & 0.551 \pm 0.073 \\
	3 & 60.0 \pm 8.2 & 3067 \pm 113 & 107.8 \pm 4.4 & 904 \pm 107 & 383 \pm 29 & 1.98 & 0.505 \pm 0.050 \\
	5 & 58.1 \pm 7.1 & 3854 \pm 153 & 103.9 \pm 4.5 & 385 \pm 143 & 416 \pm 78 & 3.55 & 1.25 \pm 0.17 \\
	7 & 58.6 \pm 7.4 & 2659 \pm 62 & 108.6 \pm 2.7 & 432 \pm 58 & 375 \pm 33 & 4.94 & 1.51 \pm 0.12 \\
	8 & 60.4 \pm 8.6 & 3265 \pm 74 & 108.1 \pm 2.8 & 551 \pm 67 & 401 \pm 33 & 5.35 & 1.65 \pm 0.15
	\enddata
\end{deluxetable}

\newpage
\clearpage

\begin{deluxetable}{CCCCCCCC}
	\tablewidth{0pt}
	\tablecolumns{8}
	\tablecaption{\label{table:table7} Same as Table~\ref{table:table6}, but for $E_{\mathrm{e}}=475$~eV.}
	\tablehead{
		\colhead{Beam current} & 
		\colhead{$\Gamma_{\mathrm{e}}$} & 
		\colhead{$A_1$} &
		\colhead{$\Gamma_1$} & 
		\colhead{$A_2$} & 
		\colhead{$\Gamma_2$} & 
		\colhead{$\bar{n}_{\mathrm{e}}$} & 
		\colhead{$n_{\mathrm{eff}}$} \\
		\colhead{(mA)} & 
		\colhead{$\mathrm{(\mu m)}$} & 
		\colhead{(counts)} & 
		\colhead{$\mathrm{(\mu m)}$} & 
		\colhead{(counts)} & 
		\colhead{$\mathrm{(\mu m)}$} & 
		\colhead{$(10^{11}\: \mathrm{cm^{-3}})$} & 
		\colhead{$(10^{11}\: \mathrm{cm^{-3}})$}
	}
	\startdata
	1 & 51.2 \pm 6.3 & 4578 \pm 152 & 94.3 \pm 3.7 & 537 \pm 121 & 414 \pm 91 & 0.082 & 0.254 \pm 0.033\\
	2 & 53.3 \pm 7.2  & 970 \pm 174 & 123 \pm 13 & 169 \pm 123 & 336 \pm 107 & 1.51 & 0.348 \pm 0.094 \\
	3 & 56.7 \pm 6.7 & 2797 \pm 447 & 103 \pm 11 & 751 \pm 263 & 267 \pm 68 & 2.02 & 0.63 \pm 0.14 \\
	5 & 57.0 \pm 6.3 & 2337 \pm 150 & 102.1 \pm 7.2 & 496 \pm 141 & 357 \pm 11 & 3.35 & 0.98 \pm 0.14 \\
	7 & 54.4 \pm 6.1 & 1663 \pm 78 & 110.4 \pm 5.4 & 476 \pm 72 & 374 \pm 38 & 5.18 & 1.13 \pm 0.12 \\
	9 & 57.4 \pm 6.5 & 1883 \pm 73 & 107.7 \pm 5.1 & 419 \pm 60 & 462 \pm 50 & 6.04 & 1.46 \pm 0.16
	\enddata
\end{deluxetable}

\clearpage

\bibliography{EBIT}

\end{document}